\documentclass[10pt]{iopart}

\usepackage{iopams}  
\usepackage{cite}  
\usepackage{graphicx}
\usepackage[breaklinks=true,colorlinks=true,linkcolor=blue,urlcolor=blue,
citecolor=blue]{hyperref}
\newcommand{\text}[1]{\textnormal{#1}}
\renewcommand{\vec}[1]{\bi{#1}}
\usepackage[utf8]{inputenc}
\usepackage[T1]{fontenc}
\usepackage{times}
\usepackage[american]{babel}

\usepackage{braket}
\usepackage{siunitx}

\newcommand{\al}{\ensuremath{\alpha}}
\newcommand{\be}{\ensuremath{\beta}}

\begin{document}

\title{Extending 
first principle
plasma-surface simulations to experimentally 
relevant scales}

\author{M Bonitz$^1$, A Filinov$^{1,2,3}$, J W Abraham$^1$ and  D Loffhagen$^2$}
\address{$^1$ Institut f\"ur Theoretische Physik und Astrophysik,
Christian-Albrechts-Universit\"at, Leibnizstr. 15, D-24098 Kiel, Germany}
\address{$^2$ INP Greifswald e.V., Felix-Hausdorff-Str.~2, D-17489 Greifswald, 
Germany}
\address{$^3$ Joint Institute for High Temperatures RAS, Izhorskaya Str.~13, 
125412 Moscow, Russia}
\ead{bonitz@theo-physik.uni-kiel.de}


\begin{abstract}
The physical processes at the interface of a low-temperature plasma and a solid are extremely complex. They 
involve a huge number of elementary processes in the plasma, in the solid as 
well as charge, momentum and energy transfer across the interface. In the majority of plasma simulations these surface processes are either neglected or treated via phenomenological parameters such as sticking coefficients, sputter rates or secondary electron emission coefficients. However, those parameters are known only in some cases, so 
such an approach is very inaccurate and does not have predictive capability. Therefore, improvements are highly needed.
In this paper we briefly summarize relevant theoretical methods from solid state and surface physics that are able to contribute to an improved simulation of plasma-surface interaction in the near future.

Even though the (quantum-mechanical) equations of motion for the participating charged and neutral 
particles are known, in principle, full \textit{ab initio} quantum simulations are 
feasible only for extremely short times and/or small system sizes. A substantial simplification is achieved when electronic quantum effects are not treated explicitly. Then one arrives at much simpler semi-classical 
molecular dynamics (MD) simulations for the heavy particles that have become the main workhorse in 
surface science simulations. Using microscopically (i.e., density functional theory) founded potentials and force fields as an input, these MD simulations approach the quality of \textit{ab initio} 
simulations, in many cases. 
 However, despite their simplified nature, these simulations require a time step  that is of the order or below one femtosecond making it prohibitive to reach 
experimentally relevant scales of seconds or minutes and system sizes of micrometers.

To bridge this gap in length and time scales without compromising the first principles character and predictive power of the simulations, many physical and computational strategies have been put forward in surface science. This paper presents a brief overview on different methods and their underlying physical ideas, 
and we compare their strengths and weaknesses. Finally, we discuss their 
potential relevance for future plasma-surface simulations.
The first class are ``acceleration'' techniques that include 
metadynamics, hyperdynamics, temperature accelerated dynamics, collective 
variable driven hyperdynamics and others. Recently we have presented a novel 
approach: \textit{Selective process acceleration} [Abraham \textit{et al.}, J. 
Appl. Phys. \textbf{119}, 185301 (2016)] which we discuss in some more detail. The second promising route to longer accurate simulations is  \textit{Dynamical freeze out of 
dominant modes} which we have introduced recently for the simulation of neutral atom sticking on a metal surface [Filinov \textit{et al.}, this issue]. In this article we give a more general view on this method that allows to accurately combine first principles MD simulations with semi-analytical models and discuss possible applications that are of potential relevance for plasma physics.

\end{abstract}

\noindent{\it Keywords}: IOP journals

\submitto{\PSST}                            

\maketitle

\ioptwocol

\section{Introduction}\label{s:intro}
%

Recent progress in low-temperature plasma physics, both, in experiments and 
applications \cite{plasma-road-map-17, doe-report-17}, 
creates an urgent need for accurate simulations of the plasma-solid interface. 
Even though there have been remarkable recent advances, both, in plasma modeling 
and surface science simulations, the combination of the two is still at an early 
stage. 
Current simulations in low-temperature plasma physics, often omit plasma-surface processes or treat them phenomenologically. Let us take as an example the treatment of neutrals. In advanced kinetic simulations based on the Boltzmann equation, e.g. \cite{hagelaar_2005,donko_2016} or particle in cell (PIC-MCC) simulations, e.g. \cite{ebert_2016,becker_2017} neutrals are treated as a homogeneous background, and their interaction with surfaces is not included in the description. However, the effect of energetic neutrals maybe crucial for secondary electron emission (SEE), as was demonstrated in PIC simulations of Derszi \textit{et al.} where neutrals above a threshold energy of 23eV were traced \cite{derszi_2015}. 
The second example is the impact of the properties of the surface, such as surface roughness or oxidation or coverage by an adsorbate, on the behavior of the plasma. Using realistic surface properties--as they emerge upon contact with a plasma--drastically alters the plasma-surface interaction compared to the case of an ideal (i.e. clean and perfect) surface. This has been studied in great detail for the case of SEE by Phelps and Petrovic \cite{phelps_1999}, and this was taken into account in PIC simulations via modified cross sections in Ref.~\cite{derszi_2015}. In this work it was found that a realistic (``dirty'' \cite{phelps_1999}) surface gives rise to a significant increase of the ion density, even far away from the electrode. The data of Ref.~\cite{phelps_1999} suggest that there remain substantial uncertainties in the values of the SEE coefficient. In a real plasma treatment experiment a ``clean'' surface may correspond to the initial state of an electrode which, ultimately, turns into a ``dirty'' metal that is covered by adsorbates or an oxide layer.

Similarly, Li \textit{et al.} studied the effect of surface roughness on the field emission by including a phenomenological geometric enhancement factor \cite{li_2013}.
The third example is related to plasma electrons hitting a solid surface. The standard assumption in simulations is that these electrons are lost without reflection, e.g. \cite{sheehan_2013}, and only recently a microscopic calculation of the electron sticking coefficient was performed by Bronold and Fehske \cite{bronold_prl15}. They also studied the charge transfer when a  strontium ion from the plasma approaches a gold surface \cite{pamperin_prb15}.

  \begin{figure}
  \begin{center} 
  \hspace{-0.cm}\includegraphics[width=0.5\textwidth]{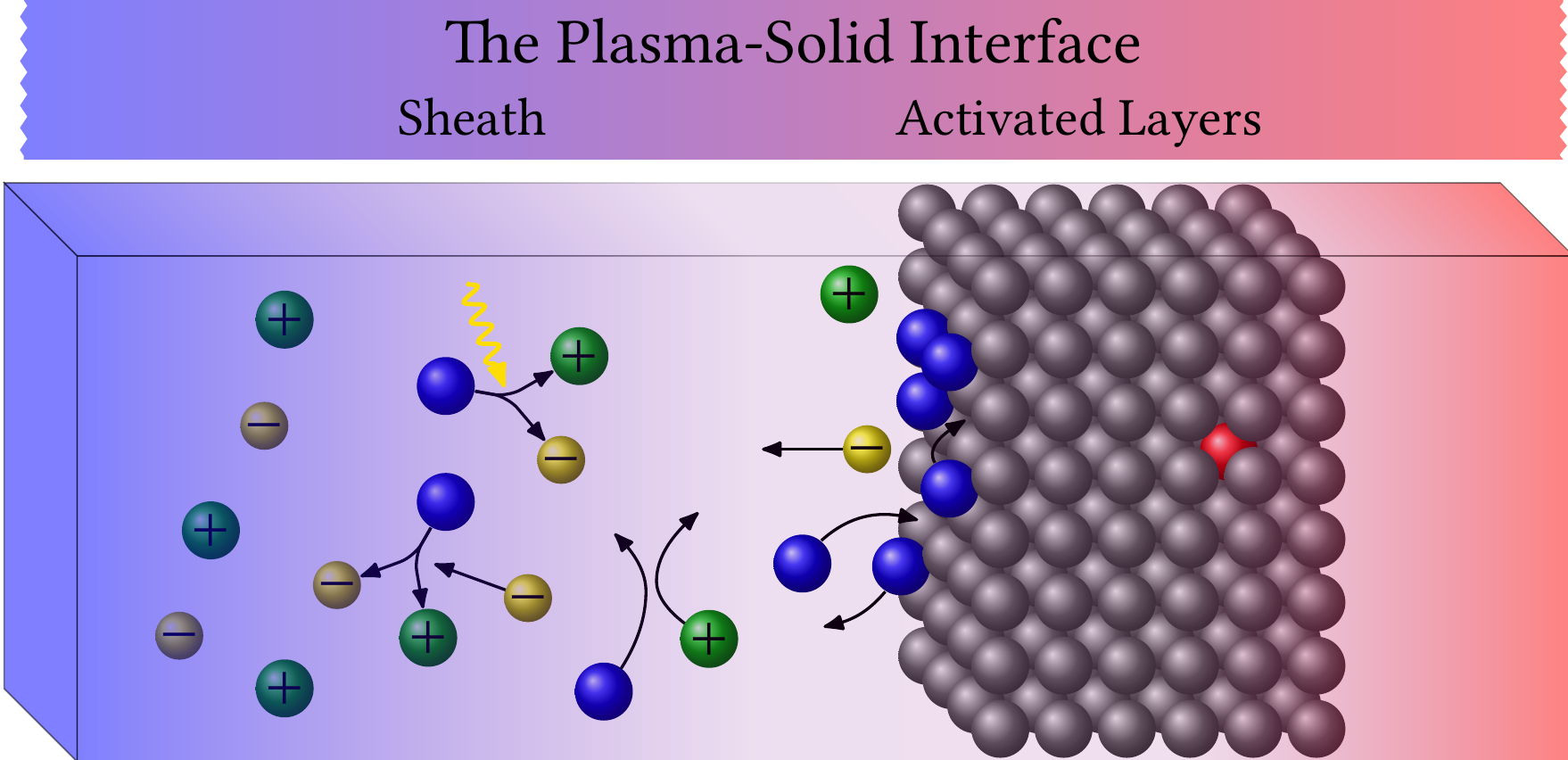} 
  \end{center}
  \caption{Sketch of the plasma-solid interface which comprises the plasma sheath and plasma facing layers of the solid \cite{interface}. Among the relevant processes are diffusion, adsorption (``sticking'') and desorption of neutrals, penetration (stopping) of ions and electron transfer between solid and plasma. Typically, in plasma simulations the effect of the surface is described by empirical parameters such as the SEE coefficient, sticking coefficients, sputter rates etc. The mutual influence of the plasma on the solid and vice versa is a major challenge for a predictive theoretical treatment and require a combination of various theoretical approaches, see Fig.~\ref{fig:theory}.}
  \label{fig:psi}
  \end{figure}
\begin{figure*}
  \begin{center} 
  \hspace{-0.cm}\includegraphics[width=0.78\textwidth]{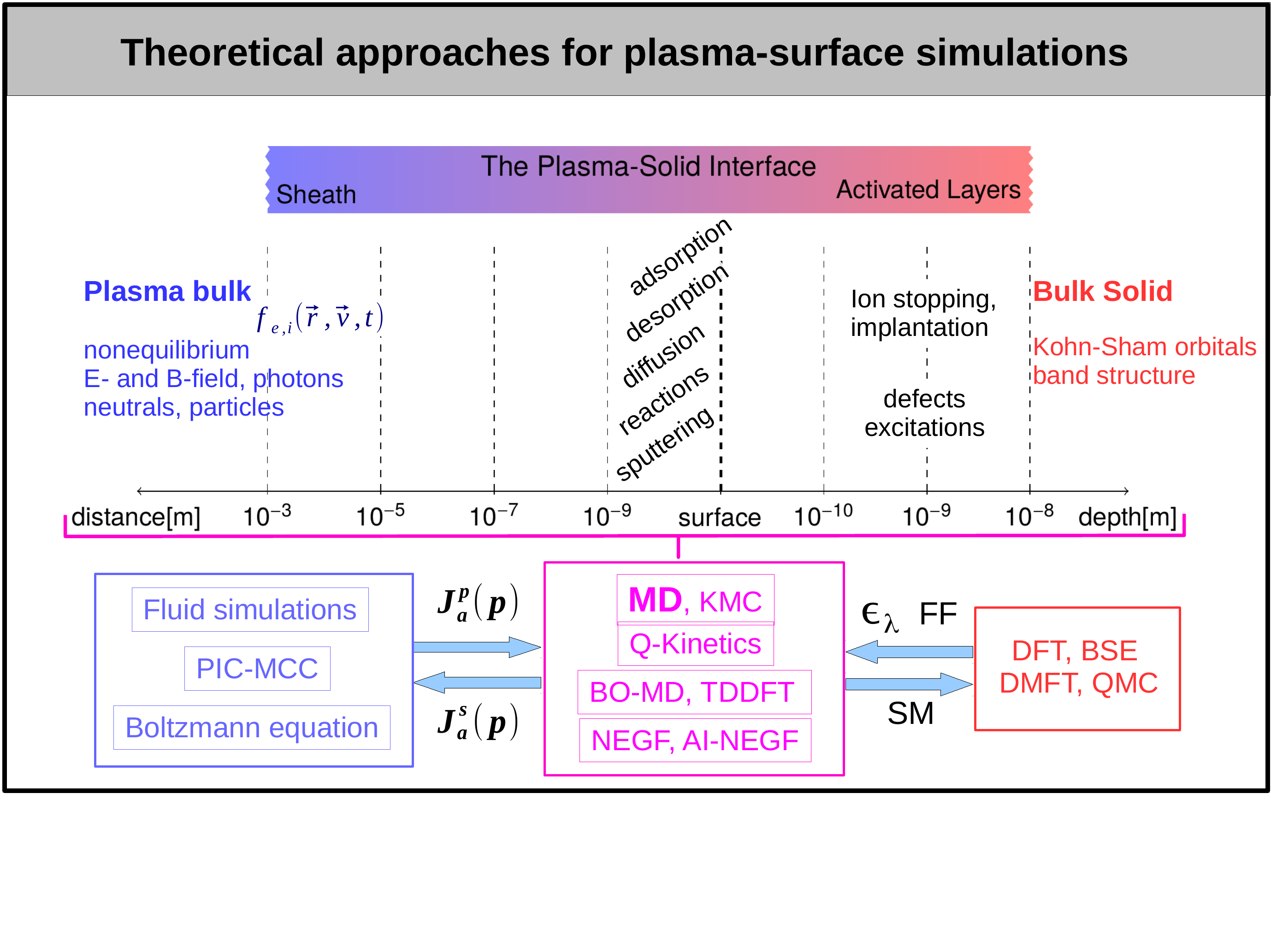} 
  \end{center}
  \vspace{-2.0cm}
  \caption{Theoretical methods for the description of the plasma-solid interface \cite{interface}, as sketched in  Fig.~\ref{fig:psi}. Some of the processes of interest are listed in the figure. Note the dramatically different length scales and the very different properties of plasma and solid requiring totally different methods to be applied on the plasma and the solid side. Standard methods for the bulk solid are Density functional theory (DFT),  Bethe-Salpeter equation (BSE), Dynamical Mean Field Theory (DMFT), and Quantum Monte Carlo (QMC). To simulate  surface processes (central box), additional non-adiabatic (time-dependent) approaches are required: molecular dynamics (MD), Kinetic Monte Carlo (KMC), Quantum Kinetics, Born-Oppenheimer MD (BO-MD), time-dependent DFT (TDDFT), Nonequilibrium Green functions (NEGF) and \textit{ab initio} NEGF (AI-NEGF). To account for the complex interactions between plasma and solid, the corresponding methods have to be properly linked: plasma simulations should provide the momentum dependent fluxes $\textbf{J}^p_a$ of all species ``a'' to the surface whereas surface simulations deliver the corresponding fluxes $\textbf{J}^s_a$ that leave the surface. Bulk solid simulations provide the band structure $\epsilon_\lambda$ and reactive force fields (FF), whereas surface simulations return the updated surface morphology ``SM'', chemical modifications etc. For details see Sec.~\ref{s:challenges}. This paper focuses on the MD approach and on the question how to increase its efficiency and provides examples for the fluxes of atoms $\textbf{J}^s_A$ (Sec.~\ref{ss:md-re}) and time-dependent surface morphology (Sec.~\ref{ss:spa}).}
  \label{fig:theory}
  \end{figure*}

 
The latter quantum-mechanics based approaches are very promising but they are still at a very early stage of development. While they clearly indicate the importance of an accurate treatment of plasma-surface processes, they cannot yet make reliable predictions.
The reason is that a huge variety of complex physical and chemical processes 
occur at the plasma-solid interface,  
which include secondary electron emission, sputtering, neutralization and 
stopping of ions, adsorption and desorption of neutral particles as well as chemical reactions, for an illustration, see Fig.~\ref{fig:psi}. Moreover, the typical particle densities in the plasma and 
the solid differ by many orders of magnitude and to completely different physics active on both sides of the interface: low-density gas-like behavior in the plasma versus quantum dynamics of electrons in the solid. Furthermore, the density gap gives rise to a huge gap in 
relevant space and time scales, cf.~Fig.\ref{fig:theory}.

\begin{description}
  \item[The first step]
 to tackle these problems is to have a look at the theoretical approaches that have been developed in solid state physics to describe a surface that is exposed to a plasma. These methods are based on density functional theory (DFT) and various additional many-body methods that allow to treat correlated materials. However, these methods typically focus on the ground state properties of the solid. In contrast, in the presence of a plasma, particle and energy fluxes to and from the solid occur giving rise to 
nonequilibrium effects and high excitation. Therefore,
\item[ as the second step,]
 one has to consider nonequilibrium methods that describe the solid and the plasma-solid interface under these conditions. These include time-dependent \textit{ab initio} (quantum) methods density functional theory, nonequilibrium Green functions, and quantum kinetic theory. However, these approaches are extremely time consuming and allow one to cover only small systems for a few femtoseconds. Therefore,  
\item[the third step] consists in additional simplifications, mostly, in eliminating the quantum effects from the dynamics of the interface. This leads to semi-classical molecular dynamics simulations for the heavy particles where all quantum effects are being ``absorbed'' into effective pair interaction potentials or force fields. With accurate force fields (typically based on \textit{ab initio} DFT simulations) the resulting MD simulations are very accurate and of first principle character (fully solving Newton's equations). This is still very challenging computationally because stable solution of these equations requires a time step of about one femtosecond. Therefore, there is no straight way to reach experimentally relevant time and length scales, even on supercomputers. This leads to 
\item[step four:]
 invoking additional physical ideas that allow one to either \textit{accelerate} or \textit{extend} these first principles MD simulations, without compromising the accuracy,  to the time scales of interest. Even though this may seem impossible, a number of powerful and successful concepts have been developed in Statistical physics, many-body physics, quantum chemistry and surface science. One of the goals of this paper is to present an overview on those concepts that might be of relevance for plasma-surface simulations in the near future.
\end{description}

This paper is organized as follows. In Sec.~\ref{s:challenges} we give a brief summary on the theoretical methods that are required to accurately simulate plasma-surface processes and discuss their problems.
In Sec.~\ref{s:concepts} we give a brief overview on the acceleration approaches that are of potential relevance 
for plasma-surface interaction. The first group of methods---acceleration of 
phase space sampling---is discussed in Sec.~\ref{s:phase_space} whereas the 
second method---coarse graining approaches---is the context of 
Sec.~\ref{s:coarse_graining}. Then we 
discuss in more detail one of the latter approaches---Dynamical freeze out of 
dominant modes---in Sec.~\ref{s:freezout}.
The conclusions are given in Sec.~\ref{s:conclusion1}.
\section{Challenges in the simulation of plasma-solid interaction}\label{s:challenges}
An accurate simulation of plasma-surface processes, first of all, requires a reliable description of the solid (step one above). To this one first of all needs to obtain the ground state properties of the solid--the energy spectrum (band structure) and the Kohn-Sham orbitals--which is done by density functional theory (DFT) simulations, cf. right part of Fig.~\ref{fig:theory}. However, DFT is known to have problems, in particular, in treating materials with strong electronic correlations including various oxides. Here, many-body approaches are being used that include the Bethe-Salpeter equation (BSE), dynamical mean field theory (DMFT) of quantum Monte Carlo (QMC).

If the solid comes in contact with a low-temperature plasma (step two above), energetic electrons and ions may excite the electrons of the solid and the lattice. This is already not captured by ground state DFT but requires time-dependent extensions, cf. the approaches listed in the central box of Fig.~\ref{fig:theory}.
Recently some elementary processes such as the impact of ions (stopping power) and neutralization of ions at a surface and chemical reactions were studied by
 \textit{ab initio} quantum simulations. This includes  Born-Oppenheimer MD (coupling of density functional theory for the electrons to MD for the ions) and time-dependent DFT (TDDFT), e.g. 
\cite{brenig-pehlke_08,zhao_prb_15}.  
 \textit{Ab initio}  nonequilibrium Green functions (NEGF) simulations 
 are an alternative  
 that allow for a more accurate treatment of electronic correlations  
\cite{balzer_prb_16, schluenzen_cpp_16, balzer_prl_18}. For completeness, we also mention \textit{ab initio} NEGF--a recent combination of ground state DFT and NEGF \cite{marini}.
However, TDDFT, NEGF and AI-NEGF simulations are extremely CPU-time demanding and can treat only small systems 
for short time scales. 
 For example, Born-Oppenheimer  MD simulation requires a 
time step around 0.1fs, which allows to treat on the order $100\dots 
1000$ atoms for $1\dots 100$ picoseconds, during a week of simulations on 
massively parallel hardware, e.g. \cite{hutter_12}. The demand for TDDFT and 
NEGF is several orders of magnitude larger.
 
 At the same time for many processes an explicit quantum modeling of the 
electron dynamics is not necessary. This concerns, in particular, the dynamics 
of neutral particles on a surface: diffusion, adsorption and desorption or many 
chemical reactions. Here, often a semi-classical MD  simulation is performed (step three above)--a technique that is well developed in surface science and in theoretical 
chemistry, e.g. \cite{gross-book}. Similarly, MD simulations are well established in low-temperature plasmas, e.g. to compute first principle structural properties of dust particles \cite{com-plasmas_14} or the diffusion coefficient in a strongly correlated magnetized plasma \cite{ott_prl_11}.
In each case, the quality of the MD results depends on the accuracy of effective pair potentials or force 
fields that are usually derived from microscopic quantum simulations or are adjusted to 
reproduce experimental data. These MD simulations are not \textit{ab initio} anymore (they neglect quantum effects in the dynamics), but still carry first principle character (they solve Newton's equations exactly), so they will be referred to as first principle MD simulations below. Typically they require a time step 
of the order of $1$ fs and can treat huge systems. For example, Ref.~\cite{nakano_08} reported simulations of a system containing $10^{11}$ 
atoms that reach times of the order of several milliseconds. However, this is 
presently only possible on the largest supercomputers or on dedicated hardware, 
e.g.~\cite{piana_13}.

Despite these impressive records, it is clear that in the near future MD 
simulations for
plasma-surface processes will remain many orders of magnitude short of system 
sizes and length scales needed to compare with experiments. 
In plasma physics, these are minutes and (at least) micrometers, respectively. Therefore, 
additional strategies are needed. One way is  of course the use 
of additional approximations leading to simplified models at the expense of 
accuracy and reliability. Here, we discuss another approach, which aims at 
retaining the 
first principles character
of the MD simulations (step four above). The idea is to 
invoke additional information on the system properties that allow one to 
effectively accelerate the simulations and/or to extend them to larger scales 
\textit{without loosing accuracy}.

There exists a variety of acceleration strategies including hyperdynamics 
\cite{voter_97}, metadynamics \cite{parrinello_pnas_02} or temperature 
accelerated dynamics \cite{Sorensen2000}. A more recent concept is collective 
variable driven hyperdynamics \cite{Bal2015} that was reported to achieve, for 
some applications, speed-ups of the order of nine orders of magnitude. Another 
approach developed by the present authors \cite{abraham_jap_16, abraham_cpp_18} 
uses a \textit{selective acceleration of some relevant processes} and also 
achieved speed-ups exceeding a factor $10^9$.
Another direction of developments does not aim at accelerating the \textit{ab 
initio} simulations but to extend them to longer times by a suitable combination 
with analytical models \cite{franke_prb_10, paper1}. These methods will be called below 
\textit{Dynamical freeze out of dominant modes} (DFDM). The goal of this article is to 
present an overview on these very diverse acceleration/extension developments, 
to discuss their respective strengths and limitations and  to outline future 
improvements and extensions for applications in plasma-surface interaction.

  \begin{figure}
  \begin{center} 
  \hspace{-0.cm}\includegraphics[width=0.52\textwidth]{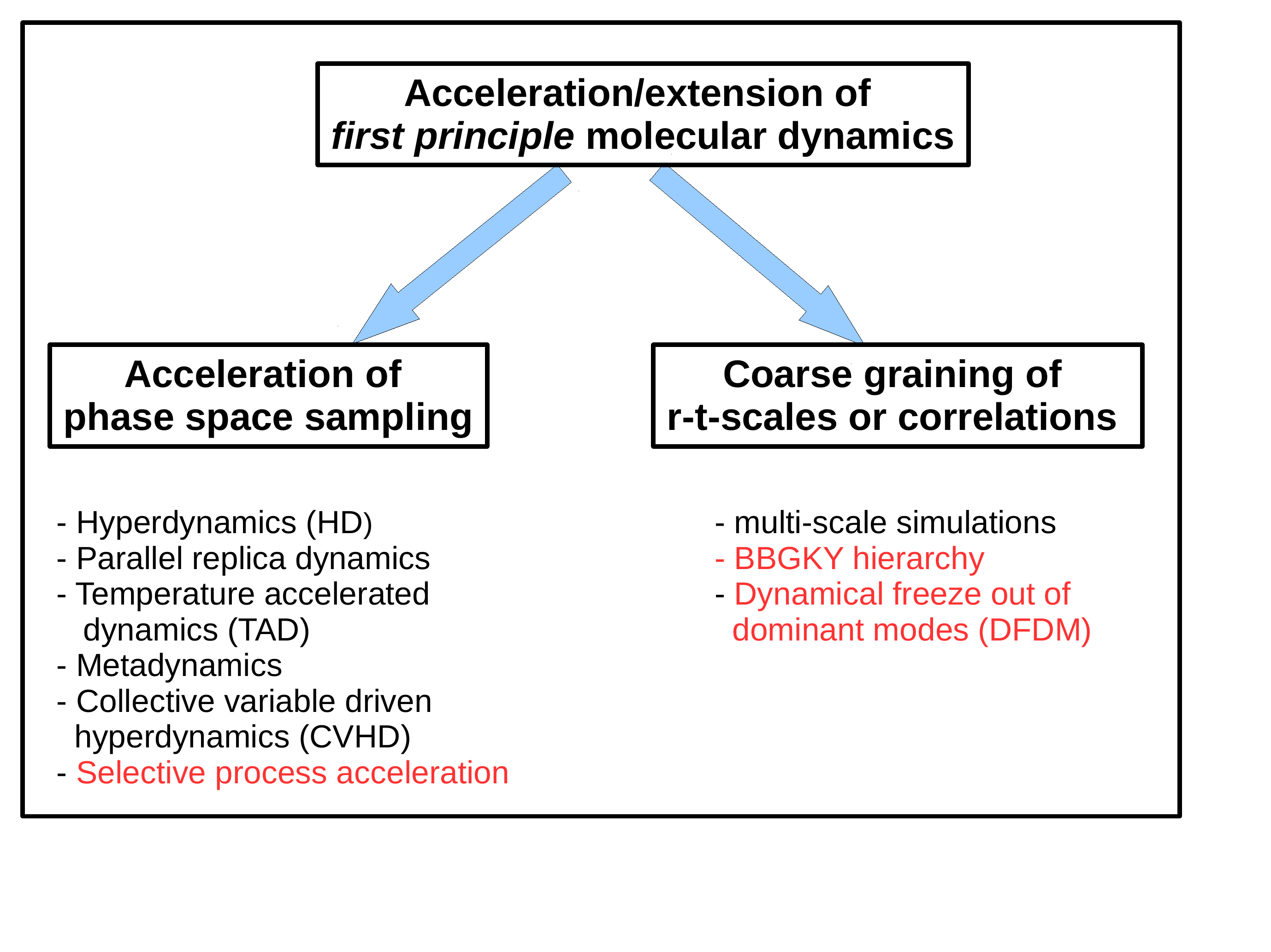} 
  \end{center}
  \vspace{-.80cm}
  \caption{Main potential strategies to accelerate/extend 
  first principles (semiclassical) MD  
simulations for plasma-surface applications. This method is primarily applicable to the dynamics of neutral particles. The left column contains common 
approaches in surface science MD. The right column sketches strategies that are 
motivated by many-body theory and plasma physics. The red items are discussed in 
detail in this paper.}
  \label{fig:scheme}
  \end{figure} 

\section{Concepts to accelerate and/or extend \textit{ab initio} MD 
simulations}\label{s:concepts}
As discussed in the introduction, 
first principles MD is based on the use of 
accurate pair potentials or force fields. The steepness of these force fields 
leads to a rather small time step of the order of one femtosecond that has to be 
used to achieve convergent simulations. As a consequence, the total simulation 
duration is far away from experimentally relevant times of seconds and even 
minutes and, therefore, acceleration strategies are of high interest. This 
problem is not specific to plasma-surface interaction, but also occurs in the 
study of phase transitions, the chemistry of macromolecules, biology, in surface 
physics and surface chemistry.

These different and diverse communities developed a large number of strategies 
to accelerate MD simulations, to improve the treatment of rare events or to 
extrapolate to longer times or larger systems. These strategies can be loosely 
grouped into two classes, which are depicted in 
Fig.~\ref{fig:scheme}. The first group (left column) includes methods that 
accelerate MD simulations by overcoming bottlenecks, such as rare events or 
trajectories being trapped in local potential minima. The second group of 
methods has been termed ``coarse graining'' approaches (right column in 
Fig.~\ref{fig:scheme}). Here the idea is to average over fast processes or small 
length scales that are not of interest for the physical observables. This is 
complementary to the first group and promising for plasma-surface simulations. 


\section{Acceleration of phase space sampling}\label{s:phase_space}
We start by discussing the concepts listed in the left column of Fig.~\ref{fig:scheme}. 
The approaches discussed in this section are used to treat systems
which reside in local energy minima for a long time before
any event of interest occurs. 
Metadynamics is represented in section~\ref{ss:metadyn}. 
The methods presented in sections~\ref{ss:hd}--\ref{ss:ParRep} --
hyperdynamics, temperature accelerated dynamics and parallel replica dynamics --
have been proposed by Voter and co-workers. 
They have in common that they aim at an effective reduction of the waiting time
between successive infrequent events.
A different approach  has been introduced in Ref.~\cite{abraham_jap_16}. 
It 
relies on treating the diffusive motion of atoms
on surfaces exclusively with Langevin dynamics.
This method  allows for a selective process acceleration and  
is discussed in Sec.~\ref{ss:spa}.
In the following, we give a brief overview of these methods;
for further details, we refer to recent reviews 
\cite{abraham_epjd_17,neyts2017molecular,Neyts2012mdmc,Perez2009_voter,Voter2002
}.

\subsection{Metadynamics}\label{ss:metadyn}
Metadynamics is a technique to enhance the computation
of a multidimensional free energy surface (FES)
of a many-body system. In most cases the FES is far too complicated to be directly computed. Laio and Parrinello
in 2002 introduced a method that allows for an efficient computation of the FES by means of molecular dynamics~\cite{parrinello_pnas_02} which has become a widely used 
tool in computational (bio-)physics,
chemistry and material science. 
An overview is given e.g.\ in Ref.~\cite{laio_ropp_08}.

The first key idea (and fundamental assumption) of the method is that the free energy $F$ of a system with a set of coordinates $\mathbf{x}$,
the potential $ V(\mathbf{x})$
and the inverse temperature  $\beta = 1 / (k_\mathrm{B}T)$
can be expressed as a
function of just
a few collective variables $\mathbf{S} = ( S_1, \ldots, S_d)$
according to
\begin{equation}
    F = - \frac{1}{\beta} \ln \left( \int \exp\left[-\beta V(\mathbf{x}) \right]
    \delta[\mathbf{S}-\mathbf{S}(\mathbf{x})] \, \mathrm{d}\mathbf{x}
    \right) \,.
\end{equation}
If the collective variables provide an adequate
representation of the whole configuration space,
the FES can be efficiently explored
by performing MD simulation where a second key idea is applied: a history-dependent small ``bias potential'' $\Delta V$ is added, i.e. $V \to V + \Delta V$. This potential successively enforces the system to leave every occurring minimum in the FES, thus, avoiding bottlenecks due to rare events.
In a very simple fashion, this bias potential can be constructed as a sum of 
weighted Gaussian functions,
\begin{equation}
    \Delta V(\mathbf{S},t)=w \sum_{t'\in \mathcal{T}} \exp \left(-\sum_{i=1}^d 
\frac{[S_i(t)-S_i(t')]^2}{2(\Delta s_i)^2} \right) \,,
\end{equation}
where the set $\mathcal{T}$ comprises all times $t'$ before the time $t$, 
at which the sum of Gaussian functions has been extended by one term.
The interval between successive creations of new Gaussian functions  
as well as their weights $w$ and widths $\Delta s_i$ 
should be chosen such that a compromise between computation time and accuracy is achieved.
After a sufficiently long simulation time the potential landscape $V+\Delta V$
levels out and becomes flat. This can be recognized in the simulation by the ``diffusive'' behavior
of the considered collective variables \cite{laio_ropp_08}.
Then the inverse of the final bias potential $\Delta V$ provides an accurate
estimator of the free energy $F$.

In contrast to the methods discussed below, the use of 
metadynamics alone does not yield correct state-to-state dynamics.
Nevertheless, it can be utilized for that purpose if
it is combined with other methods,  such as the collective variable driven 
hyperdynamics (CVHD)  described in  section~\ref{ss:hd}.
Metadynamics is a versatile method and it allows
for a relatively easy implementation. 
Hiwever, the choice of the collective variables 
can be very difficult.

\subsection{Hyperdynamics}\label{ss:hd}
Using hyperdynamics \cite{Voter1997}, the state-to-state dynamics
of an infrequent event system is accelerated
by 
adding a space-dependent bias potential $\Delta V(\mathbf{r})$ to the potential 
energy surface $V(\mathbf{r})$. Thereby, the energy barriers between different 
states are reduced
so that transitions occur more often.
For the applicability of the method,
it is required that
both the unbiased and the biased system dynamics obey
the so-called transition state theory (TST) 
\cite{Marcelin1915_tst,Eyring1935_TST}. 
Furthermore, the bias potential $\Delta V(\mathbf{r})$ must be zero at all 
dividing
surfaces, and it must be chosen such that
the correct relative probabilities of the transitions are maintained.

The construction of an appropriate bias potential  can be an elaborate task in 
many cases. 
In the original publication~\cite{Voter1997},
the diffusion of an Ag$_{10}$ cluster on an Ag(111) surface
was investigated by constructing
$\Delta V$ as a function of the lowest eigenvalue
of the Hessian matrix.
At this, boost factors of roughly \SI{8e3}{} were achieved.

Another approach introduced by Fichthorn \textit{et al.}~\cite{Fichthorn2009},
the so-called bond-boost method, is to let $\Delta V$ be a function of the
nearest-neighbor bond lengths in a solid.
In a study of the diffusion of Cu atoms on a Cu(001) surface
in the temperature range between \SI{230}{K} and \SI{600}{K},
boost factors of up to $10^6$ could be achieved \cite{Miron2003bondboost}.

Even higher boost factors of up to $10^9$ were obtained for the same system
using the CVHD method 
introduced by Bal and Neyts \cite{Bal2015}. The idea of this method 
is to use the concept of metadynamics for an incremental build-up
of a bias potential depending on just one collective variable, i.e., a variable 
that describes the relevant processes in the system.
The CVHD method can be applied whenever the requirements of hyperdynamics
are fulfilled and an appropriate collective
variable can be found.
Even though the latter may be difficult in some cases, the CVHD method
has the potential to be applied to many kinds of different systems.
For example, it has already been used to study 
the folding of a polymer chain model \cite{Bal2015}
and the pyrolysis and oxidation of $n$-dodecane \cite{Bal2016fuel}.

\subsection{Temperature accelerated dynamics}\label{ss:tad}
The idea of temperature accelerated dynamics (TAD) \cite{Sorensen2000} is to 
make transitions occur more often by performing
the simulation 
at an elevated temperature
$T_\mathrm{high}$ instead of the temperature of interest $T_\mathrm{low}$. 
Because this procedure alone would induce wrong
ratios of escape probabilities, 
an additional mechanism is applied to correct for this.
The method can only be applied if the system obeys the harmonic TST (HTST). 
Thus, 
it is more restrictive than 
hyperdynamics and parallel replica dynamics (section~\ref{ss:ParRep}), for which
the harmonic approximation is not necessary.

TAD is carried out by performing ``basin constrained
molecular dynamics'' (BCMD) for each system state. 
Whenever a transition occurs at $T_\mathrm{high}$,
the escape path and the corresponding escape
time are stored, but
the involved particles are reflected back to their initial
energy basin and the dynamics is continued.
For each observed transition at time $t_\mathrm{high}$,
the transition time is extrapolated
to the lower temperature according to
\begin{equation}
    t_\mathrm{low} = t_\mathrm{high} \exp \left\{ E_\mathrm{a} 
\left(\frac{1}{k_\mathrm{B}T_\mathrm{low}}
    - \frac{1}{k_\mathrm{B}T_\mathrm{high}} \right) \right \}\,,
\end{equation}
where $E_\mathrm{a}$ is the energy barrier of the transition.
The BCMD routine is stopped when
the simulation time reaches
\begin{equation}
    \tilde{t}_\mathrm{high} = \frac{\ln(1/\delta)}{\nu_\mathrm{min}}
    \left(
    \frac{\nu_\mathrm{min}t_\mathrm{low,min}}{\ln(1/\delta)}
    \right)^{T_\mathrm{low} / T_\mathrm{high}}\,,
\end{equation}
where $t_\mathrm{low,min}$ is the minimum of the extrapolated
 transition times at $T_\mathrm{low}$,
$\nu_\mathrm{min}$ is a guess for a lower bound of the pre-exponential
factors occurring in the formulas for all possible transitions,
and $\delta$ is 
a pre-defined limit for the probability to observe every transition after 
$\tilde{t}_\mathrm{high}$
that would replace the transition at the current minimum $t_\mathrm{low,min}$.
As a result of the procedure, the simulation time is advanced by 
$t_\mathrm{low,min}$,
and the process corresponding to this time is executed.

The boost factor that can be achieved by means of TAD  depends critically 
on the ratio of $T_\mathrm{high}$ to $T_\mathrm{low}$. As one cannot choose 
arbitrarily
high values of  $T_\mathrm{high}$ without breaking the requirements of HTST,
the method is particularly effective for systems at low temperature.
For example, a boost factor of $10^7$ was achieved
in a simulation of the growth of a Cu(100) surface at $T_\mathrm{low} = 
\SI{77}{K}$ \cite{Voter2002}.
A recent example for a simulation at a higher temperature of $T_\mathrm{low} = 
\SI{500}{K}$ 
can be found in Ref.~\cite{Georgieva2011tad}, where the
sputter
deposition of Mg--Al--O films was studied.

\subsection{Parallel replica dynamics}\label{ss:ParRep}
While standard parallelization techniques
are usually applied to extend the accessible system sizes,
parallel replica dynamics (ParRep) allows one to
use parallel computing to 
extend the time scales, too \cite{Voter1998}.
Among the three methods hyperdynamics, TAD and ParRep,
ParRep is the most accurate one, and a higher boost can be trivially
achieved by increasing the number of processors $N_\mathrm{p}$ 
\cite{Perez2009_voter,Neyts2012mdmc}.
ParRep can be applied to any infrequent event system with first-order
kinetics, i.\,e., with exponentially distributed
first-escape times
of all occurring processes, 
\begin{equation}
f(t) = \lambda \exp \left(-\lambda t \right) \,. 
\end{equation}

The ParRep procedure starts by replicating and
dephasing the system on each available processor.
Each copy of the system is
propagated   independently and in parallel on each processor 
until a transition is detected
on one of the processors.
Then, the system clock is  advanced by the sum of the 
$N_\mathrm{p}$ individual simulation times, and the global
system state is set to the state reached after the observed transition.
Subsequently, a short serial run is performed to allow for the occurrence
of correlated events. After that, the whole procedure is repeated.

As the number
of processors is 
one limiting factor of the 
achievable speed-up, the ParRep  method is often less effective than
hyperdynamics and TAD. Nevertheless, it is simple to implement, 
and it can be combined with other acceleration methods.
Therefore, ParRep has  become
a valuable tool in the field of computational materials science 
\cite{Perez2015parrep}.
For example, it has been applied to simulate
the diffusion of H$_2$ in crystalline C$_{60}$ \cite{Uberuaga2003ex_parrep},
the diffusion of lithium in amorphous polyethylene oxide 
\cite{Duan2005ex_parrep}, and the
crack-tip behaviour in metals \cite{Warner2007ex_parrep}.

\subsection{Selective process acceleration (SPA). MD simulation of gold film growth on a polymer surface}\label{ss:spa}
\begin{figure}[h]
  \begin{center} 
  \hspace{-0.2cm}\includegraphics[width=0.49\textwidth]{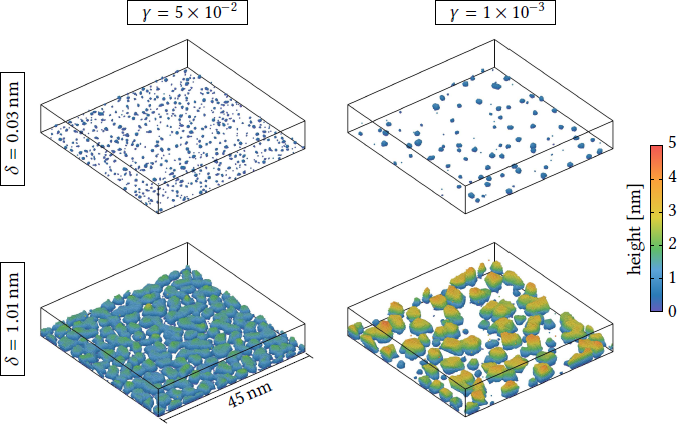} 
  \end{center}
  \caption{Time evolution of the gold film morphology deposited on a polystyrene substrate from accelerated MD simulations. Figures show the 3D-cluster configuration in real space. Top row: early time corresponding to a film thickness of $0.03$nm. Bottom row: later time, see also Fig.~\ref{fig:abraham}. The influence of the plasma is varied from the left to the right column. Right: defect fraction due to energetic ions, $\gamma=0.001$. Left: $\gamma=0.05$.
  From Ref.~\cite{abraham_diss_18}. 
}
  \label{fig:abraham-snap}
  \end{figure}

In some cases, it is reasonable to assume that
the motion of atoms on a surface or in a medium
is approximately Brownian.
This type of motion can be generated by 
solely performing Langevin dynamics for the particles of interest, while  
the other atoms and molecules of the background medium
do not have to be explicitly included in the simulations.
Here we consider, as an example, the deposition of gold atoms onto a polymer surface. The MD simulations tracked each individual atom, its diffusion on the surface, the emergence and growth of clusters and, eventually the coalescence of the latter. A typical example is presented in Fig.~\ref{fig:abraham-snap} and shows the cluster configuration at an early moment (top) and a later time point (bottom). 

The influence of the plasma environment is mostly due to the impact of energetic ions. This leads to the formation of surface defects that trap incoming atoms and prevent their diffusion.
The figure compares the cases of a weak plasma effect (right column, the fraction of atoms trapped equals $\gamma=0.001$) and a stronger effect (left column, $\gamma=0.05$.)
In the former case, a small number of large clusters is being formed, due to cluster coalescence, whereas in the latter case the film is much more homogeneous, containing a much larger number of smaller clusters \cite{abraham_diss_18}. 

We underline that the lower snapshots in the figure refer to a film thickness of about one nanometer which requires a deposition time of about two minutes. This is impossible to achieve with first principle MD simulations. The key to achieve this extreme simulation duration and to compare to experiments was selective process acceleration (SPA). The main ideas of the approach are explained in the following.

In the MD simulations the  isotropic Langevin equation of motion for all gold particles
with the mass $m$ and spatial coordinates $\mathbf{r}=(\mathbf{r}_1, 
\mathbf{r}_2, \ldots)$
are solved:
\begin{equation}
m \ddot{\mathbf{r}} = - \nabla U(\mathbf{r})
- \frac{m}{t_\mathrm{damp}} \dot{\mathbf{r}}
+  \sqrt{\frac{2m k_\mathrm{B}T}{t_\mathrm{damp}}} \mathbf{R}\,,
\label{eq:eqmotion_langevin}
\end{equation}
where the potential $U$ describes the interaction between gold particles. For this potential \textit{ab initio} force field data are being used (the MD simulations used the LAMMPS package).
Further, 
$t_\mathrm{damp}$ has the role of a damping parameter,
and
$\mathbf{R}$
is a delta-correlated Gaussian random process.
This random force and the viscous damping simulate the effect of the polymer on the heavy gold particles.
If only neighborless atoms are considered, i.\,e., $\nabla U \equiv 0 $,
the combination of the last two terms on the right side of
Eq.~(\ref{eq:eqmotion_langevin})
induces a diffusive motion with the diffusion coefficient
\begin{equation}
\label{eq:diffusioncoeff}
D=\frac{1}{m} {k_\mathrm{B}T t_\mathrm{damp}}\,.
\end{equation}
Thus, it is  clear that the utilization of the Langevin dynamics allows one to control
the speed of the surface diffusion and bulk by choosing a specific combination of the temperature 
$T$
and the damping parameter $t_\mathrm{damp}$.
An anisotropic diffusive motion
can be generated if
one generalizes Eq.~(\ref{eq:eqmotion_langevin}) by
separately defining damping parameters $t_\mathrm{damp}^x$, $t_\mathrm{damp}^y$
and $t_\mathrm{damp}^z$ for each of the three spatial directions.
Beyond that, it is possible to add a spatial dependence to the diffusion 
coefficient
if one lets the damping parameters depend on the position of the particle. 

Based on the above considerations,
Abraham \textit{et al}.~\cite{abraham_jap_16} developed a procedure to
simulate the growth of nanogranular gold structures on a thin polymer film.
Instead of simulating the polymer with explicit particle models,
their method relies on performing Langevin dynamics for the gold atoms
with the simulation box being partitioned into three parts
representing the upper part of the polymer bulk (I), the surface of the polymer
(II) and the region above the surface (III). 
By choosing appropriate ratios of the damping parameters,
one can make sure that the atoms spend most of the time in the surface layer 
(II), where they
perform a random walk which is restricted to a small range of possible 
$z$-coordinates.
The use of Langevin dynamics is restricted to regions (I) and (II);
in region (III), the dynamics is purely microscopic. 
This allows one to add particles to the system by creating particles
at the top of the simulation box and assigning them a negative initial velocity.
Therefore, it is  possible to perform the simulation with values of the 
deposition rates $J_\mathrm{sim}$
and diffusion coefficients $D_\mathrm{sim}$ that are much higher than the values 
in typical
experiments.

In Ref.~\cite{abraham_jap_16}, it was argued that the simulations
yield an adequate description of a real experimental deposition process
if the ratio $J_\mathrm{sim}/D_\mathrm{sim}$ is equal
to the ratio 
$J_\mathrm{exp}/D_\mathrm{exp}$ 
of the corresponding quantities in the experiment.
The idea behind that is that -- at least at the early stage of the
deposition process -- the growth should be essentially
determined by the average distance an atom travels on the surface between 
successive
depositions of atoms. 
Hence, the absolute time of the process is  assumed to be irrelevant.
  \begin{figure}[h]
  \begin{center} 
  \hspace{-0.2cm}\includegraphics[width=0.5\textwidth]{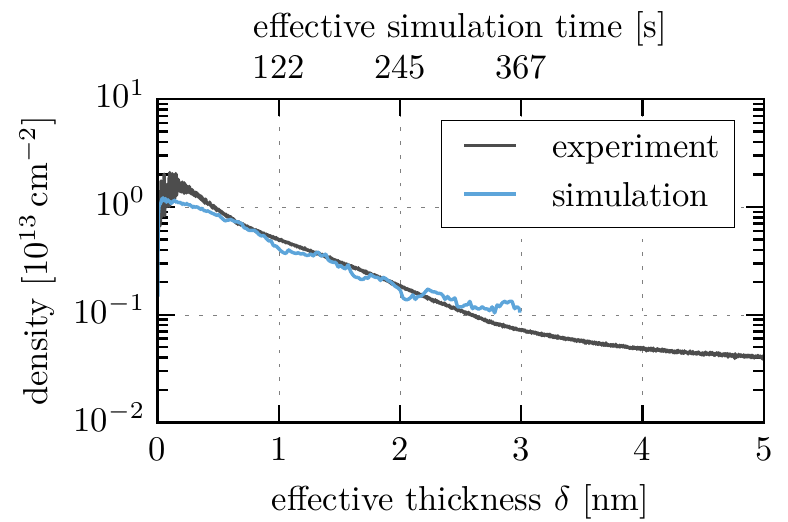} 
  \end{center}
  \caption{Number density of gold clusters on a polymer film as a function
  of the effective film thickness.
  The data has been taken from Ref.~\cite{abraham_jap_16} 
  where results of MD simulations with SPA 
  are compared with data
  from GISAXS experiments~Ref.~\cite{Schwartzkopf2015}.
  The upper horizontal axis of the plot shows
  the impressiv effective simulation time reached by \textit{accelerating 
  the deposition of atoms and the diffusion} of atoms on the surface. 
  }
  \label{fig:abraham}
  \end{figure} 
The results presented in Ref.~\cite{abraham_jap_16}
were obtained with a time step of $\SI{1}{fs}$,
and a damping parameter for the diffusion in $x$- and $y$-directions of 
$\SI{1}{ps}$. 
The temperature and the deposition rate were set to match the conditions of the 
experimental results in Ref.~\cite{Schwartzkopf2015} for the sputter deposition 
of gold on polystyrene.
Using these parameters, the direct MD simulation time for the growth of a thin 
gold film
is roughly $10^9$ times shorter than the corresponding time in the experiment. Or in other words, the duration of the MD simulations could be extended by nine orders of magnitude.
To verify the validity of such a dramatic shift of the time scales,
comprehensive tests of the method were performed, see also 
Refs.~\cite{abraham_diss_18,abraham_epjd_17}
for a discussion.
In particular, as one  accelerates only selected processes, i.e., the deposition 
of atoms 
and the diffusion of atoms on the surface,
one has to make sure that the neglect of other processes,
e.\,g., the relaxation of a cluster structure,
does not lead to artifacts in the simulation results. 
In Ref.~\cite{abraham_jap_16}, 
the method was tested by comparing
several quantities describing the evolution of the gold film morphology
with the results of time-resolved in situ grazing incidence x-ray scattering (GISAXS) experiments of Schwartzkopf \textit{et al.} \cite{Schwartzkopf2015}. It turned out that many of the experimentally
observed features could be reproduced for film thicknesses
up to \SI{3}{nm}. This thickness corresponds to an impressive effective 
simulation
time of \SI{367}{s} which is directly suited for comparison with measurements. 
As an example of the compared quantities,  
figure~\ref{fig:abraham} shows the number density of metal clusters on the 
polymer surface. The comparison 
between experimental data and simulation results 
shows very good agreement, at least up to a time of about \SI{350}{s}. For 
longer times, the simulation start to deviate from the measurements indicating that the  procedure is no longer applicable.

The present approach of selective acceleration of dominant processes can be 
generalized to other systems as well. 
A recent application concerned 
the deposition and growth of bi-metallic clusters on a polymer surface 
\cite{abraham_cpp_18} where the acceleration allowed one to study the very slow 
process of demixing of the two metals. Applications of this approach to various plasma processes should be possible as well. One effect that has already been studied 
is the creation of defects by ion impact. The main effect is trapping of 
clusters \cite{abraham_jap_15} 
at the defect locations which reduces their diffusion and limits cluster coalescence, see Fig.~\ref{fig:abraham-snap} above.
In addition to the deposition of neutral atoms, 
the method also allows one to describe the impact of ions and the growth of 
charged clusters.

In concluding this section we mention that similar problems of rare events 
appear not only in MD simulations but also in statistical approaches such as 
Monte Carlo simulations. Some strategies are discussed in Ref.
\cite{boening_prl_08}, where  further references are given as well.

\begin{table*}{\Large $\;\qquad$ {\bf  Hierarchy of scales and relaxation 
processes in many-body systems}}\\[3ex]
\begin{tabular}{|c|c|c|c|}
\hline
&&&\\
Time and  & \textbf{Stage}, Effects & Quantities & Theory \\
length scales &&&\\ [2ex]
\hline \hline
&&&\\
\textbf{IV} & {\bf Equilibrium} & $n^{\rm EQ}_a$, $T^{\rm EQ}$, $p^{\rm EQ}$ &
Equilibrium theory \\
&& $p=p(n_1, n_2 \dots,T)$ & Equation of state\\
$t > t_{\rm hyd}$&& $n_a=n_a(n_1,n_2\dots,T)$ & Mass action law \\
& Correlated equilibrium, or& $p=p^{\rm ideal}+p^{\rm cor},\;$ etc.&\\[2ex]
$l > l_{\rm hyd}$ & Stationary nonequilibrium state & $n_a({\cal U})$, $T({\cal 
U})$, $p({\cal U})$ &
Quasi-equilibrium theory \\ 
&in an external field ${\cal U}(\textbf{R})$ &&\\[2ex]
\hline
&&&\\
\textbf{III} & {\bf Hydrodynamic Stage}
& $n_a({\bf R}t)$, $\textbf{u}_a({\bf R}t)$, $T_a({\bf R}t) $ &
Hydrodynamic equations \\[2ex]
$t\in[t_{\rm rel},t_{\rm hyd}]$ & Local equilibrium & $f_a=f^{\rm EQ}_a 
\Big(n({\bf R}t),\textbf{u}({\bf R}t),T({\bf R}t)\Big)$ & Gas-dynamic 
equations\\
$l\in[l_{\rm mfp},l_{\rm hyd}]$ &
& 
& Reaction-diffusion eqs. \\
& Correlation corrections & $\:n_a({\bf R}t) = n_a^{\rm ideal}({\bf R}t) + 
n_a^{\rm cor}({\bf R}t)$  & Rate equations\\
&& etc. & Master equation\\
[2ex]
\hline
&&&\\
\textbf{II} & {\bf Kinetic Stage} & $ f_a({\bf p}{\bf R},t) $
& Kinetic theory/ \\
&  & $f_a(t_0)$& Relaxation time\\
$t\in[t_{\rm cor},t_{\rm rel}]$ &  Functional hypothesis && approximation \\
& Equilibrium correlations&  $g_{ab}=g^{\rm EQ}_{ab}(\{f(t)\})$ & Markov limit 
(M) +\\
$l\in[l_{\rm cor}, l_{\rm mfp}]$ & Kinetic energy conservation&$\;\,\quad = 
g^{\rm M}_{ab,0}+g^{\rm M}_{ab,1}+ \dots$
& Correlation corrections\\
[2ex]
\hline
&&&\\
\textbf{I} & {\bf Initial Stage}
& $g_{ab}({\bf p}_a{\bf R}_a{\bf p}_b{\bf R}_b,t)$
& Generalized  \\
&Initial correlations & $g_{ab}(t_0)$ &kinetic equations \\
$t\in[t_0,\tau_{\rm cor}]$&Correlation buildup & & Correlation time approx.\\
&Total energy conservation &&  \\
$l < l_{\rm cor}$ &Higher correlations & $g_{abc}$,  $g_{abcd}, \dots$ & 
first principle simulations 
\\
[2ex]
\hline
\end{tabular}\\[2ex]
\caption{Characteristic scales and relaxation processes in correlated many--particle systems (schematic). Typical examples are the relaxation of electrons in a plasma following local ionization or excitation by a short electric field pulse or the thermalization of atoms from the plasma on a solid surface.
Beginning at the initial time $t_0$, the evolution goes (from bottom to top) 
through several time stages and extends from small to larger length scales. This 
can be viewed as successive \textit{coarse graining}, cf. Fig.~\ref{fig:scheme}. 
Accordingly,
the relevant observables and
concepts for a statistical description change.  For explanations and details, 
see Sec.~\ref{s:coarse_graining}. Adapted from Ref.~\cite{bonitz_qkt}. }
\label{tab1}
\end{table*}

\section{Coarse graining approaches}\label{s:coarse_graining}
\subsection{General idea}
We now discuss the approaches listed in the right column of 
Fig.~\ref{fig:scheme}. The main idea 
of the coarse graining approaches 
is to perform an analysis of the different length and time scales that exist in 
a many-particle system -- such as the plasma-surface system -- driven out of 
equilibrium [one example could be a dense system of ions that is excited by an electric field pulse. Another example could be an ensemble of neutrals from the plasma that impact a solid surface and equilibrate there due to collisions with the lattice atoms].
Depending on the type of excitation and on the properties of the system, 
the relaxation towards equilibrium 
typically proceeds 
in a number of steps. Even though this is a highly complex process 
in general, 
it is often possible to identify a sequence of relaxation processes or even a 
hierarchy. 

A situation 
typical of gases and plasmas is sketched in table~\ref{tab1}. Here, 
four relaxation stages  are distinguished which are separated by the following 
characteristic time scales: 
the hydrodynamic time scale $t_{\rm hyd}$, as well as the kinetic time scales  
$t_{\rm rel}$ (relaxation time) and 
$\tau_{\rm cor}$ (correlation time). The latter is directly related to an 
equilibration of pair, $g_{ab}$ (triple, $g_{abc}$, and higher) correlations, 
$t_{\rm rel}$ denotes the time necessary to establish a Maxwell (equilibrium) 
distribution $f^{\rm EQ}$, and $t_{\rm hyd}$ is associated with the decay of 
density, velocity or temperature fluctuations or inhomogeneities or similar 
large-scale excitations. At each of these stages the system is adequately 
described by a specific set of quantities, and their dynamics is governed by 
specific equations: hydrodynamic equations (Stage III), kinetic equations (Stage 
II) and generalized non-Markovian kinetic equations or \textit{ab initio} 
simulations (Stage I), respectively. 

Even though each of these equations is an approximation to the full many-body 
equations, 
 these equations are accurate within their respective stages and time scales. 
 Thereby, the accuracy of these models is typically not limited by the equations 
themselves (e.g. resulting from the omission of higher order terms) but by the 
parameters entering these models. 
 For example,  these parameters are the transport or hydrodynamic coefficients 
(determined by the distribution function) 
 in  hydrodynamic equations. 
 In the case of kinetic equations, these parameters are cross sections or 
collision integrals. 
 Of course,  approximation schemes are used in practice  
 for these parameters. But here we consider a different (although hypothetical) 
situation: \textit{if these input quantities would be known exactly, the 
underlying model equations would be exact}, within their respective range of 
applicability\cite{comment_exactness}. Of course, this requires the application 
of rigorous coarse graining procedures for the derivation of these equations, 
some examples and properties of which we will discuss in 
Secs.~\ref{ss:averaging}, \ref{ss:environment} and \ref{ss:bbgky}.

The existence of formally exact coarse grained equations is at the heart of this 
work and 
is 
explored in the remainder of this paper. In particular, our main idea is to 
obtain these exact input quantities for models from 
first principle MD 
simulations and thereby realize our goal to significantly extend the duration of 
first principle
simulations. Since the coarse grained equations (Stages 
II-IV) emerge dynamically in the course of the equilibration, this novel method 
is called \textit{Dynamical freeze out of dominant modes} (DFDM). We  
demonstrate it for simple examples in Sec.~\ref{s:freezout}. 

\subsection{Time dependencies during equilibration}
An  observation 
of interest for the following discussion is that the distribution function 
$f_a(\textbf{R},\textbf{p},t)$ 
[$a$ could denote ions or neutrals, in the examples above] is still time- and space-dependent, 
even after its equilibration. However, this dependence is only implicit 
and arises exclusively from the (slower) time evolution and weaker space 
dependence of the macroscopic fields entering the function 
\begin{equation}
    f_a(\textbf{R},\textbf{p},t) = f_a^{\rm 
EQ}[n(\textbf{R},t),\textbf{u}(\textbf{R},t), T(\textbf{R},t)],
\end{equation}
i.e., the density $n$, the mean velocity $\textbf{u}$ and the mean energy 
(temperature $T$). 
By contrast, the momentum dependence is fixed by the Maxwellian form. 
This applies to the hydrodynamic Stage (III), cf.~Tab.~\ref{tab1}, whereas the 
distribution carries, in addition, an explicit time dependence, on the kinetic stage (II). 
This concept is based on the \textit{functional hypothesis} of Bogolyubov 
\cite{bogoljubov}. 
Close to the border of the two stages, i.e. for $t$ slightly below $t_{\rm 
rel}$, the deviations of $f$ from 
 the distribution function 
$f^{\rm EQ}$ at equilibrium are small, and the time derivative of $f$ 
[the collision integral, see Eq.~(\ref{eq:col-integral})] 
can be approximated by the linear relation
\begin{equation}
I_f(t) \approx   - \frac{1}{t_{\rm rel}} \left[ f(t)-f^{\rm EQ} \right].
    \label{eq:rta}
\end{equation}
This is nothing but the familiar \textit{relaxation time approximation} (or 
Bhatnagar–Gross–Krook (BGK) 
collision integral \cite{bgk-integral}), where the relaxation time is usually 
taken from equilibrium models of the system which have limited accuracy. In 
cases where \textit{ab initio} simulation data are available, as we assume here, 
the MD 
simulation 
results can be used to extract an \textit{ab initio} relaxation time which 
transforms Eq.~(\ref{eq:rta}) into an exact relation. This also requires to use 
the MD data for the equilibrium distribution that may be modified in the 
presence of correlations between the particles. This improved description  also 
allows for a more accurate modeling of the hydrodynamic stage.
We mention 
for the sake of completeness,
that a similar crossover can also be studied between stages I and II. 
In that case, the pair correlation function reaches its equilibrium form $g^{\rm 
EQ}$ 
around the correlation time  $t\sim \tau_{\rm cor}$ 
and retains only an implicit time dependence via the distribution functions 
$g(t)=g^{\rm EQ}[f(t)]$ (cf.~Tab.~\ref{tab1})  
thereafter. 
Similar to the relaxation time approximation, here one can use a 
\textit{correlation time approximation} to describe the final approach to the 
equilibrium correlations \cite{bonitz_pla_96,bonitz_qkt}.

\subsection{Averaging over time and/or length scales}\label{ss:averaging}
As a simple example, 
we consider the 
second order differential equation 
\begin{eqnarray}
 \frac{\mathrm{d}^2{\hat A}}{\mathrm{d}t^2} + \gamma \frac{\mathrm{d}{\hat 
A}}{\mathrm{d}t} + {\hat k}(t){\hat A} = 0\, ,
\label{eq:fast_oscill}
\end{eqnarray}
which  represents e.g.\ a Newtonian equation of motion. 
Here, $\gamma$ is a dissipation coefficient, and ${\hat k}(t)$ denotes a quickly 
fluctuating force constant or frequency. 
This could be a dust particle in a complex plasma that experiences collisions with plasma neutrals,  ions or electrons. Another example could be a large molecule on a solid surface that undergoes collisions with the lattice atoms.
A frequent situation is that ${\hat k}(t)=k+\delta \hat{k}$, where $k\equiv 
\langle {\hat k}\rangle$, and the brackets denote time averaging over the period 
of the fast oscillations or rapid dynamics of the light particles. Correspondingly, the dynamics of the variable $\hat A$ 
can be split into a slowly changing term and a rapidly oscillating contribution 
according to ${\hat A}(t)=A+\delta \hat{A}$,
where $A=\langle {\hat A}\rangle$ obeys the  equation of motion
\begin{eqnarray}
 \frac{\mathrm{d}^2 A}{\mathrm{d}t^2} + \gamma \frac{\mathrm{d} A}{\mathrm{d}t} 
+ k(t)A = - \langle \delta{\hat k}(t) \delta{\hat A(t)} \rangle = I_A\,.
\label{eq:slow_oscill}
\end{eqnarray}
This equation describes the ``coarse grained dynamics'' where the fast ``random 
fluctuations'' seem to be eliminated, and indeed, the left-hand side of this 
equation coincides with the original equation~(\ref{eq:fast_oscill}). However, 
the fast processes leave a trace on the slow dynamics via the new term on the 
right side, which is the correlation function of two fluctuating quantities. 
If the 
right-hand side is known, equation~(\ref{eq:slow_oscill}) will still be exact. 
In fact, it is not difficult to derive the equation of motion for $\langle 
\delta{\hat k}(t) \delta{\hat A(t)} \rangle$ by first finding the equation for 
$\delta{\hat A}(t)$, cf. e.g. \cite{klimontovich_sp, bonitz_un}. However, it is 
easy to see that this equation is not closed as well, and couples to a new 
quantity, $\langle \delta{\hat k}(t)\delta{\hat k}(t) \delta{\hat A(t)} 
\rangle$, which obeys its own equation of motion. Thus, an infinite
hierarchy of equations emerges which is the direct consequence of the averaging 
procedure.

The common solution is to decouple this hierarchy by invoking additional 
assumptions on the fast dynamics. A common approximation is to assume 
$\tau^{\delta {\hat k}}_{cor} / t_{A} \to 0$ which means that the correlation 
time of the rapid process (i.e. of the light particles) is vanishingly small compared to the characteristic 
time scale of $A$ (the heavy particle). In that case, the term on the right side of Eq.~(\ref{eq:slow_oscill}) 
becomes a delta correlated random process, and one recovers a Langevin-type 
equation with a Gaussian white noise term.

The above model is representative for a large variety of problems containing  
multiple time scales. A similar situation appears in the case of multiple length 
scales, where a \textit{spatial averaging} leads to a coarse grained 
description. 

A further example for a coarse grained description 
is \textit{many-body dynamics in phase space}, which is described by a 
generalized distribution function ${\hat N}(\textbf{r},\textbf{p},t)$ -- the 
microscopic phase space density  introduced by Klimontovich 
\cite{klimontovich_sp}.  
This distribution function 
obeys a mean-field type equation
\begin{eqnarray}
 \left\{\frac{\partial}{\partial t} + \textbf{v}\cdot\nabla_\textbf{r} + {\hat 
\textbf{F}}\cdot \nabla_\textbf{p}
\right\}{\hat N}(\textbf{r},\textbf{p},t) = 0\, ,
\label{eq:nhat}
\end{eqnarray}
where $\hat \textbf{F}$ denotes the total force on the particles which includes 
all external and induced forces. In general, the force also contains rapid and 
slow contributions. 
Thus, we can write again $\hat \textbf{F}=\textbf{F}+\delta\hat \textbf{F}$, and 
the above coarse graining procedure can be repeated. 
In fact, this procedure leads to the well-known Boltzmann-type kinetic equations 
for the one-particle distribution function $ f(\textbf{r},\textbf{p},t) \equiv 
\langle{\hat N}(\textbf{r},\textbf{p},t)\rangle$. It has the same form as 
Eq.~(\ref{eq:nhat}) 
and reads 
\begin{eqnarray}
 \left\{\frac{\partial}{\partial t} + \textbf{v}\cdot\nabla_\textbf{r} +  
\textbf{F}\cdot \nabla_\textbf{p}
\right\}f(\textbf{r},\textbf{p},t) = I_f(\textbf{r},\textbf{p},t)\, . 
\label{eq:kin_eq}
\end{eqnarray}
However, it contains the collision integral $I_f$ on the 
right-hand side in addition, which  is directly determined by the correlation 
function  
\begin{eqnarray}
 I_f(\textbf{r},\textbf{p},t) = - \langle \delta\hat \textbf{F}\cdot 
\nabla_\textbf{p}\delta\hat N\rangle\, 
\label{eq:col-integral}
\end{eqnarray}
of the fluctuations. 
As before, the collision term is an unknown quantity, and one can derive an 
equation of motion for it. 
Again it  includes a higher-order correlation function, and the whole system 
turns into an infinite hierarchy of equations. 
The standard solution is to use physical approximations for the choice of the 
collision integral $I_f$ using e.g.\ the Boltzmann collision integral or 
improvements such as the Balescu-Lenard integral~\cite{bonitz_qkt}.

Now, let us discuss how the idea of the present paper can be applied here. In 
cases where we have 
first principle
simulation data at our disposal, there 
exists a straightforward way how the hierarchies of equations discussed above 
can be decoupled. 
When performing MD simulations of a classical system, every observable can be 
computed. 
For example, it is possible to evaluate the right-hand side of 
Eq.~(\ref{eq:slow_oscill}) 
during a MD simulation. However, the result for $\delta {\hat k}(t)\delta {\hat 
A}(t)$ is random, depending on the choice of the initial conditions for the 
particle trajectories. The simple solution consists in running many independent 
MD simulations. Thereby, 
the initial conditions 
have to 
be chosen with a proper probability distribution 
given by the ensemble over which the averaging in the collision term $I_A$ is 
performed.
Then, the 
first principle
MD 
result for $I_A$ follows by averaging over $M$ realizations according to 
\begin{eqnarray}
 I_A(t) = \lim_{M \to \infty}\frac{1}{M}\sum_{l=1}^M
\left(\delta {\hat k}(t)\delta {\hat A}(t)\right)^{(l)}\,.
\label{eq:mean_ia}
\end{eqnarray}
This method is in principle exact for a classical system. 
The efficiency crucially depends on the cost of the MD simulation and on the 
number $M$ of trajectories needed to obtain a converged result. At the same 
time, the fluctuation around the result (\ref{eq:mean_ia}) gives an estimate of 
the statistical uncertainty. A similar approach can be applied to the kinetic 
equation (\ref{eq:kin_eq}) and the collision integral $I_f$, 
Eq.~(\ref{eq:col-integral}).

Of course, the question arises what is gained by this approach compared to 
performing only MD simulations without resorting to the many-body equation 
(\ref{eq:slow_oscill}) or the kinetic equation (\ref{eq:kin_eq}) at all. The 
point is that accurate MD simulations are 
 typically 
substantially more costly than the latter approaches. Therefore, an advantageous 
compromise between accuracy and computational effort consists in performing MD 
simulations, for short time scales, and in continuing the simulation by solving a 
kinetic equation, for longer times. 
For the latter, 
the use of the MD input for the collision integral,
as 
explained by Eq.~(\ref{eq:mean_ia}), 
could yield a significant increase in accuracy and, it could eventually offer a 
way  to extend 
first principle
simulations 
to longer times. We  return to this issue in section~\ref{s:freezout}.

Notice  that similar approaches exist also for quantum systems. An example is 
the ``Stochastic Mean field'' approach, cf. Ref.~\cite{lacroix_prb_14} and 
references therein.  Instead of MD simulations, here 
time-dependent Hartree-Fock simulations are performed over which the averaging 
is carried out. The result of this procedure turned out to be very encouraging when compared to exact simulation results, 
at least for short and intermediate time scales \cite{lacroix_prb_14}.

\subsection{Averaging over ``environmental'' degrees of 
freedom}\label{ss:environment}
Another application of coarse graining concepts is frequently used for particles 
in contact with a reservoir or ``bath''. Again this can be heavy dust particles in contact with lighter plasma particles or a macromolecule in a plasma or on a surface.
The complete state of the system of $N$ 
particles and $N_B$ bath particles is described by the phase space distribution 
function $F(\textbf{R},\textbf{P};\textbf{R}_B, \textbf{P}_B)$, where we use the 
compact notation $\textbf{R}=\textbf{r}_1, \textbf{r}_2\dots \textbf{r}_N$ and 
$\textbf{P}=\textbf{p}_1, \textbf{p}_2\dots \textbf{p}_N$ for the particles and 
similar notations for the bath particles and restrict ourselves to classical 
particles. It is easy 
to formulate the equations of motion for the whole system, but the solution for 
the $N+N_B$ particles is extremely costly. More importantly, one is typically 
not interested in the details of the dynamics of the bath particles, where 
usually $N_B\gg N$.

Therefore, the standard procedure is  to switch to a ``reduced'' description, 
which resolves only the degrees of freedom of the system, by integrating over 
the bath parameters according to 
\begin{eqnarray}
 f(\textbf{R},\textbf{P}) \equiv \int d \textbf{R}_B d \textbf{P}_B 
\,F(\textbf{R},\textbf{P};\textbf{R}_B, \textbf{P}_B)\,.
\nonumber 
\end{eqnarray}
If the bath is in thermodynamic equilibrium at temperature $T_B$, this coarse 
graining procedure transforms the dynamics of the system of particles from the 
microcanonical ensemble 
into the canonical or grand-canonical ensemble. The corresponding equations of 
motion can then be solved by Langevin MD simulation or by using a Nose-Hoover 
thermostat \cite{nose}.

Here, the main assumption  is of course the equilibrium of the bath. 
This neglects the influence of the dynamics of the system particles on the bath 
particles which might be questionable, in particular at strong excitation 
conditions. In that case, an alternative strategy consists in performing 
short-time MD simulations including the bath dynamics by using different 
realizations of the bath over which an averaging is performed, in analogy to 
Eq.~(\ref{eq:mean_ia}). This accurate procedure is 
computationally 
expensive and cannot be extended to long times. 
Thus, strategies can be developed, where one switches to the equilibrium 
description of the bath at times exceeding the thermalization time $t_B^{\rm 
rel}$ of the bath. A similar procedure is used in the simulation of disordered 
systems where  an average over different realizations of the disorder 
is performed. 

Such a strategy should be applicable to certain plasma-surface simulations such 
as scattering of plasma particles from a surface or diffusion of an adsorbate atom or molecule on a surface. In 
that case it can be justified to treat the surface as a ``bath'' at sufficiently 
long time scales. In contrast, the initial time period of the interaction of a 
plasma particle with the surface requires a full dynamic treatment of the adsorbate and the 
surface atoms. A similar idea is realized in Sec.~\ref{ss:md-re}.

\subsection{Reduced distribution functions. BBGKY hierarchy}\label{ss:bbgky}
Finally, we discuss an important approach to treat correlated many-particle systems which is based on the concept of reduced distribution functions. A typical example is the dynamics of electrons following a strong excitation and their subsequent thermalization, due to electron-electron collisions.
The dynamics of a classical $N$-particle system can be treated by 
first principle
MD simulations or, equivalently, by the $N$-particle distribution function 
$F_N(x_1, x_2, \dots x_N)$ where $x_i=(\textbf{r}_i, \textbf{p}_i)$ and $F_N$ 
obeys the Liouville equation 
\begin{eqnarray}
 f_s(x_1\dots x_s) &= &\frac{N!}{(N-s)!} \nonumber \\ 
 && \times
 \int d x_{s+1}\dots d x_N\,F_N(x_1\dots x_N)\, .
 \label{eq:Liouvilleequation} 
\end{eqnarray}
The equation of motion for $f_s$ follows directly from the original equation for 
$F_N$ by integration over the remaining variables, as in the definition 
(\ref{eq:Liouvilleequation}) of $f_s$. The resulting equation for $f_s$ is not 
closed but involves contributions from $f_{s+1}$ giving rise to a hierarchy of 
equations---the BBGKY 
(Bogoliubov-Born-Green-Kirkwood-Yvon)
hierarchy, which is discussed in detail e.g.\ in Ref. \cite{bonitz_qkt}. 

If the system is non-interacting, the $s$-particle distribution 
is a product of $s$ single-particle factors, e.g.\ $f_2(x_1, x_2)=f_1(x_1) 
f_1(x_2)$. 
Here, we do not consider quantum exchange effects. 
In case of correlations between the particles, this relation is generalized to 
\begin{eqnarray}
 f_2(x_1, x_2)=f_1(x_1) f_1(x_2) + g_2(x_1, x_2)\, ,
 \label{eq:g2}
\end{eqnarray}
where $g_2$ is the pair correlation function.
The BBGKY hierarchy is decoupled by invoking 
physically motivated approximations e.g. for the pair correlation function 
\begin{eqnarray}
 g_2(x_1, x_2,t) \to g^{app}_2([f_1];x_1, x_2,t) \, ,
 \label{eq:bbgky_decoup}
\end{eqnarray}
which is a given functional of the one-particle distribution function. As a 
result one obtains a closed equation for $f_1$---the kinetic equation---where 
the pair correlation function determines the collision integral  
$I_f=I_f[g_2^{app}]$, which coincides with the previous result given by 
Eq.~(\ref{eq:col-integral}).

In the spirit of the present paper, we underline that the resulting equation for 
$f_1$ is a dramatic simplification compared to the 
description of the 
full dynamics of all $N$ particles, in particular when $N$ is macroscopically 
large. This is the result of a very efficient coarse graining procedure. 
However, 
the quality of the result depends on the accuracy of the approximation, 
$g^{app}_2([f_1];x_1, x_2,t)$. As discussed above, reliable approximations exist 
for limiting cases, e.g.\ when the problem contains small parameters (such as for weak interaction) and for 
long times $t\ge \tau_{cor}$. However, for the initial time period, 
$0\le t < \tau_{cor}$, the standard (Markovian) results for $g^{app}_2$ are known 
to fail, 
e.g.~\cite{bonitz_qkt, bonitz_pla_96}. Here,  
the concept of the present paper can be utilized again:  
perform MD simulations at short time scales, use their result to reconstruct the 
exact functions $g_2([f_1],t)$, and extend this result to longer times using the 
kinetic equation for $f_1$. 

Such a procedure has not yet been realized so far for kinetic equations. For 
this reason it is interesting to look at other examples where this concept has 
been tested.

\section{Dynamical freeze out of dominant modes (DFDM)}\label{s:freezout}
Let us now turn to the final approach listed in Fig.~\ref{fig:scheme} under 
coarse graining concepts, right column.

\subsection{Coupled DFT-master equation approach for molecule diffusion on a 
metal surface}\label{ss:dft-me}

Franke and Pehlke performed extensive 
density functional theory 
(DFT) 
simulations of the diffusion of a 1,4-butaneditiol molecule on a gold surface 
\cite{franke_prb_10}. They found the local adsorption energy minima of the 
molecule and then applied the 
nudged elastic band (NEB) approach \cite{henkelmann_jcp_00} 
to compute the transition rates between them. This allowed them to record the 
entire ``network'' of atomic scale diffusion paths of the molecule on the 
surface. 
There is a large number of processes that are analyzed in 
Ref.~\cite{franke_prb_10}. 
Here, we are not concerned with the details of the associated diffusion hops, 
Ref.~\cite{franke_prb_10}--
but focus on the coarse graining idea.

In order to connect the \textit{ab initio} short-time diffusion simulations to 
the long-time behavior 
Franke and Pehlke
considered a master equation~\cite{franke_prb_10}
\begin{eqnarray}
\frac{dp_i(t)}{dt} &=& \sum\limits_{j\ne i}
\left\{ \Gamma_{j\to i}\, p_j(t) - \Gamma_{i\to j}\, p_i(t)
\right\},\; 
\label{eq:master}   
\\
& &0 \le  p_i(t) \le  1, \quad \sum_i p_i(t)=1\,, \nonumber
\end{eqnarray}
where $i$ is a multi-index numbering the configurations of the molecule which 
have a probability $p_i(t)$, and $\Gamma_{a\to b}$ are the transition rates 
(probability per unit time) from state $a$ to $b$. The first term on the right 
side of Eq.~(\ref{eq:master}) describes processes which increase the probability 
to realize state $i$ (``gain''), whereas the second term describes the analogous 
loss processes. For the computation of the transition rates the authors used 
standard 
TST~\cite{vineyard_57},
\begin{eqnarray}
  \Gamma_{a\to b} = \nu^0_{a\to b} \, e^{-\Delta E_{a\to b}/k_BT}\, .
 \label{eq:tst}
\end{eqnarray}
Here, $\Delta E_{a\to b}$ is the energy barrier for the transition between 
states $a$ and $b$ 
which is computed using DFT, and $\nu^0$ is the attempt frequency which is of 
the order of $10^{12}$\,s$^{-1}$ \cite{franke_prb_10}.

The authors of this reference consider two stages of the evolution: the initial 
stage, corresponding to stage I in table~\ref{tab1}, and the asymptotic hydrodynamic state, 
corresponding to stage III.
In stage I the dynamics  depend strongly on the initial configuration of the 
molecule, and the diffusion retains a memory of the initial state being 
anisotropic. In contrast, 
one expects spatially isotropic motion of the molecule in stage III, where 
averaging over many initial configurations is assumed,  
since all memory of the initial state has been lost. 
Correspondingly, it is expected that the standard diffusion equation 
\begin{eqnarray}
 \frac{\partial n(\textbf{r},t)}{\partial t} = D \Delta n(\textbf{r},t)\, ,
\label{eq:diffusion}
\end{eqnarray}
holds with the well-known time-dependent solution of the initial value problem 
for the initial condition $n(\textbf{r},0) = N\delta(\textbf{r}-\textbf{r}_0)$,
\begin{eqnarray}
 n(\textbf{r},t) = \frac{N}{4\pi D t}\,e^{-(\textbf{r}-\textbf{r}_0)^2/(4Dt)}\,.
\label{eq:diff-solution}
\end{eqnarray}
The solution (\ref{eq:diff-solution}) was recovered 
in Ref.~\cite{franke_prb_10} 
by mapping of the $p_i(t)$ on the associated spatial coordinates of the center 
of mass of the molecule, giving rise to the space-dependent probability density 
$P(\textbf{r},t)$. 
$P(\textbf{r},t) = n(\textbf{r},t)/N$ is proportional to the particle density 
n(\textbf{r},t), where $N$ is the total number of molecules (number of initial 
configurations). In the long-time limit, an exponential temperature dependence 
of the diffusion coefficient 
results, i.e., an Arrhenius law 
\begin{eqnarray}
 D(T) = a_0^2\, \nu \, e^{-\Delta E/(k_BT)}\, ,
\label{eqDTArrhenius}
\end{eqnarray}
which 
made it possible 
to recover the effective attempt frequency $\nu$ and effective diffusion energy 
barrier $\Delta E$. In (\ref{eqDTArrhenius}), $a_0$ is the surface lattice 
constant.

To summarize,  \textit{ab initio} results for the elementary diffusion motions 
of a molecule on a surface have been obtained in this example. 
While it provides a complete  microscopic picture of surface diffusion, 
this information is  far too detailed for many purposes, 
in particular,  for comparisons with measurements. 
In order to characterize the mobility of the molecules, the main interest 
concerns 
the long-time behavior, which is governed by much simpler physics described by 
the classical diffusion equation (\ref{eq:diffusion}). This reduced dynamics 
\textit{emerges dynamically} during the course of the evolution of the system 
due to self-averaging effects. 
The main advantage of this approach is 
 that the involved diffusion coefficient can be obtained exactly from \textit{ab 
initio} DFT data instead of using standard approximate results from transport 
models.

At the same time the present combination of DFT and a master equation approach 
is in principle 
able to provide additional information, beyond that presented in Ref.~\cite{franke_prb_10}. First of all, it would be possible to 
establish the equilibration time scale $t_{\rm rel}$, when the system reaches 
the isotropic diffusion regime (stage III in table~\ref{tab1}). Furthermore, it 
should be possible to  investigate the transient behavior being relevant at 
shorter time (stage II) as well, and this might give rise to a modified 
diffusion equation.

Finally, we note that similar master equation based approaches have been applied 
to the computation of chemical reaction rates, cf. \cite{maranzana_pccp_07, 
robertson_pccp_07} and references therein. In the context of plasma-surface interaction, diffusion and chemical reaction rates in the presence of a plasma are of high interest. Therefore, the present approach might be useful to derive improved surface diffusion and reaction models that take into the influence of a plasma. 

\subsection{Discussion of the validity of the master 
equation}\label{ss:valid-me}
At this point,  a first assessment of the concept to couple 
first principles
data with an analytical model is given. 
The central question concerns, of course, 
the validity limits of the master equation.  
Let us summarize the corresponding requirements.
\begin{description}
  \item[i)] First, it is known that  
  TST 
  is the basis for the transition rates~(\ref{eq:tst}), and it assumes that the 
surface is in thermal equilibrium.
  \item[ii)] All energy barriers have to be large compared to the thermal 
energy. 
  \item[iii)] The transition probabilities in the master equation 
(\ref{eq:master})  depend 
  linearly on the current occupation probabilities, i.e., the rates 
$\Gamma_{a\to b}$ are independent of all $p_i(t)$.
  \item[iv)] The master equation is Markovian, i.e., the transition 
probabilities depend only on the current state of the system (no memory).
  \item[v)] The transition rates $\Gamma_{a\to b}$ in the master equation are 
time-independent. 
\end{description}
Let us  ask the question now, how important these conditions are and which of 
them can  
eventually be relaxed, if we  would have 
first principle
simulation data at our 
disposal. First of all, conditions i) and ii) can be easily dropped and the 
rates 
can be replaced by numerical simulation results, $\Gamma_{a\to b} 
\longrightarrow \Gamma^{\rm sim}_{a\to b}$. Second, the restrictions iii)--v) on 
the master equation can be dropped as well in favor of  numerical results 
$\Gamma^{\rm sim}_{a\to b}(\{p_i\},t)$, which are updated during the simulation. 
Obviously, these rates depend generally on the probabilities as well.
 
In fact, such a generalized version of the master equation which takes into account memory of the previous system states, is nothing but a 
generalized non-Markovian kinetic equation~\cite{bonitz_qkt}, which can be 
derived rigorously from the fundamental equations of many-body physics, such as 
the BBGKY-hierarchy, cf. Sec.~\ref{ss:bbgky}. 
The necessary requirements to such a generalized master equation are: 
\begin{enumerate}
    \item[A.] The microstates of the many-body system are mapped onto a complete 
set of events ``i'' with well defined probabilities, i.e. $\sum_i p_i(t)=1$, and 
$0\le p_i(t)\le 1,$ 
    for all $i$ and  all times. 
    \item[B.] There has to be a rigorous and numerically stable procedure how to 
identify these states and to assure ergodicity.
    \item[C.]  There has to be a consistent, stable and sufficiently accurate 
procedure how to determine the corresponding probabilities and transition rates 
``on the fly'' during a simulation.
\end{enumerate}
A first example how to realize such a procedure is given in the following 
section~\ref{ss:md-re}.


\subsection{Coupled molecular dynamics--rate equation approach for atom 
adsorption on a metal surface}\label{ss:md-re}
Filinov \textit{et al.} \cite{paper1, paper2}
presented a first application of the procedure outlined above to the adsorption 
dynamics and sticking probability of argon atoms on a platinum surface. They 
performed semi-classical MD simulations of the atom dynamics using \textit{ab 
initio} pair potentials. These simulations yield the complete information on the 
particle trajectories $x_i(t)=\{\textbf{r}_i(t), \textbf{p}_i(t)\}$ with 
$i=1\dots N$. These trajectories depend on the initial conditions such as the 
incident energy 
$E_i(0)$ and angle $\theta_i(0)$, and position $\textbf{r}_i(0)$. 
All observables of interest can in principle be computed 
from these trajectories (microstates) 
without resorting to additional approximations such as in TST (\ref{eq:tst}).

  \begin{figure}[h]
  \begin{center} 
  \hspace{-0.cm}\includegraphics[width=0.48\textwidth]{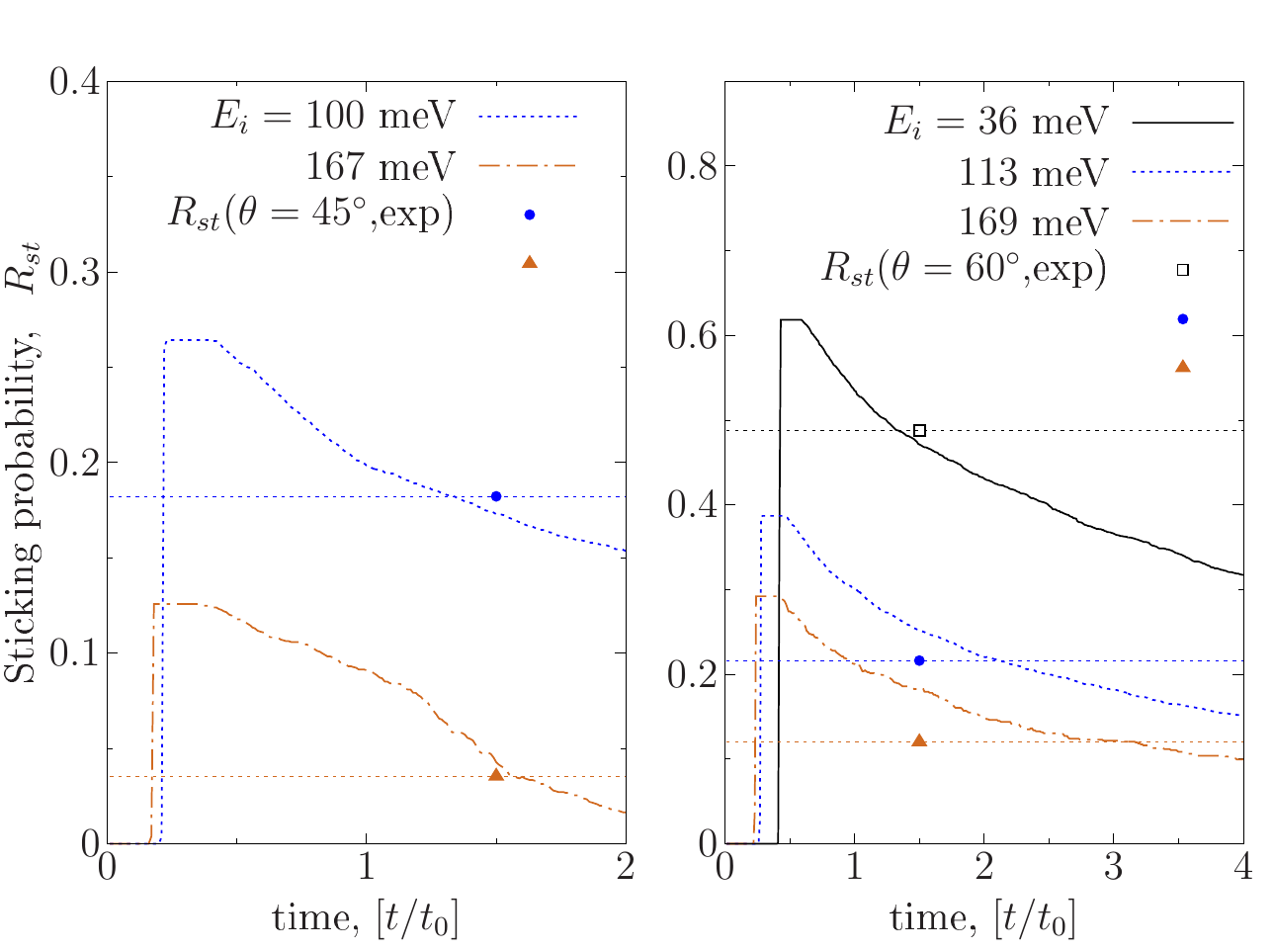} 
  \end{center}
  \vspace{-.5cm}
  \caption{Sticking coefficient of argon atoms from first principle MD simulations, as a function of time for two incident angles $\Theta$ and several impact energies $E_i$, for a lattice temperature $T_s=300K$. The experimental data (symbols \cite{expPt}) are placed at the equilibration time, $t_D=1.5t_0=10$ps obtained in the simulations. For details see Ref.~\cite{paper2}.}
  \label{fig:fig13_paper2}
  \end{figure} 
Using the procedure described below, Filinov \textit{et al.} computed the sticking coefficient of argon atoms, as a function of time and obtained very good agreement with experiments \cite{paper2}, as is illustrated in Fig.~\ref{fig:fig13_paper2}. Interestingly, the results provide energy and angle resolved data for the probability that an impacting atom will be adsorbed at the surface (or be reflected). This is valuable input information for microscopic plasma simulations. 

Let us now return to the idea of this paper--how to extend these simulations to longer times. Even though for the present problem of computing the sticking probability MD simulations of $10 \dots 40$ps duration are sufficient and no  extensions are required, it is very instructive to analyze the potential of this scheme.
First, it is clear that macroscopic properties such as the sticking 
coefficient do not require the complete microscopic information of all particle trajectories. Therefore, we attempt to map this microscopic information on a finite set of many-body states that are distinguished by the energy of the atoms.
%

The kinetic energy $E_{\text{p}}$ contains two orthogonal 
contributions and reads
\numparts
\begin{eqnarray}
 E_{\text{p}} &=& E_{\text{p}}^{\perp} + E_{\text{p}}^{\parallel}\,, \qquad
 E_{\text{p}}^{\perp, \parallel} = \frac{(p^{\perp, \parallel})^2}{2m}
 \,.
 \end{eqnarray}
 \endnumparts
In addition, each particle moves in the potential landscape of all surface atoms 
$V$ 
giving rise to the total energy 
\begin{equation}
 E_t(\vec r)=E_{\text{p}}(\vec r)+V(\vec r)\,.
 \label{eq:etot}
\end{equation}
All  trajectories of the atoms can be uniquely classified by their energy 
at every time: 1.) There are atoms with positive surface-normal energy, i.e., 
$E_{\text{p}}^{\perp} + V >0$. These particles are desorbed from the surface and 
their fraction is denoted as $N_C$ 
(continuum states). 2.) The remaining atoms have $E_{\text{p}}^{\perp} + V \le 
0$ and 
belong into two fractions. The first fraction %
has a positive total energy, 
i.e., $E_t>0$. 
These atoms can freely move across the surface and are 
denoted by $N_Q$ (fraction of ``quasi-trapped'' atoms). 
3.) The remaining atoms have a negative total energy ($E_t<0$). 
That is they are trapped in local potential minima, and their fraction is 
denoted by $N_T$. 

\begin{figure*}[h]
  \begin{center} 
  \hspace{-0.cm}\includegraphics{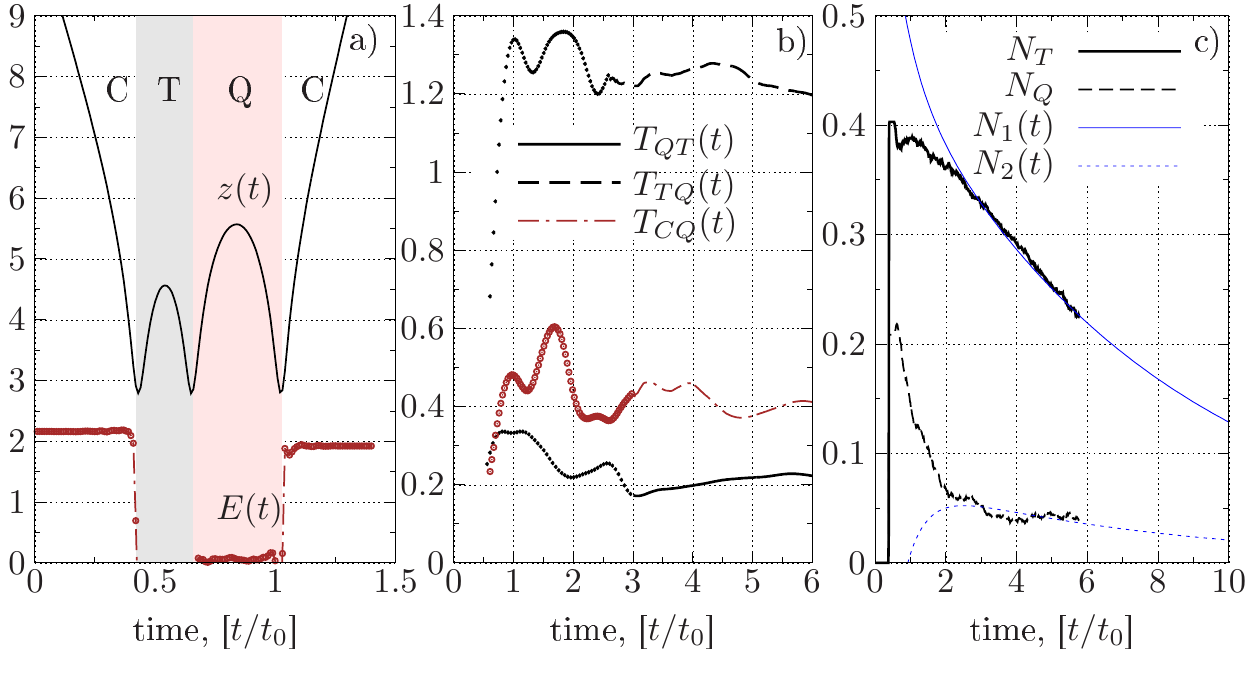} 
  \end{center}
  \vspace{-1.0cm}
  \caption{Illustration of the combined MD-rate equations approach for atom 
sticking \cite{paper1}. \textbf{a)}: Example of an Ar-atom trajectory at a 
platinum surface at temperature $T_s=300$K. $z(t)$: height above the surface (in 
Angstroms). $E(t)$: total energy of the atom (in 10 meV). Incident conditions 
(energy and angle): $E_i=21.6$~meV and $\theta_i=45^\circ$.  Depending on the 
energy the atom belongs to one of the three categories: continuum (C), 
quasi-trapped (Q) or trapped (T). \textbf{b)}: Time evolution of the three 
dominant transition rates (in units $t_0^{-1}$). The rate $T_{\al\be}$ denotes 
the transition $\beta \rightarrow \al$. The smallest rate, $T_{CT}\approx 
0.093$, is not shown.  \textbf{c)}: Fraction of trapped and quasi-trapped atoms 
as a function of time. Dashed and dotted lines: MD simulation data. $N_1$ and 
$N_2$ are the associated solutions of the rate equations using stationary 
transition rates. The time unit is $t_0=6.53$ps. 
  }
  \label{fig:alex}
  \end{figure*}

An example of such a trajectory is depicted in Fig.~\ref{fig:alex}a. There, an 
atom approaches the surface (being in a continuum state) and by colliding with the surface atoms, rapidly looses a 
large fraction of its energy, becoming 
trapped. Afterwards it  gains again energy from another collision with the surface 
(thereby being transferred to a quasi-trapped state) until it is eventually desorbed (returning to a continuum state C).

The main advantage of this mapping approach is that 
a large statistical ensemble of trajectories is available 
 for sufficiently many atoms
leading to an excellent accuracy of the results. 
We underline that this classification of all atoms into just three categories 
(three ``states'') is unique, i.e., $N_Q(t)+N_T(t)+N_C(t)=1$, in agreement with 
the  condition A  listed in section~\ref{ss:valid-me}. 
In fact, these fractions can be understood as occupation probabilities $p_i$ 
with $i=\{C, Q, T\}$ of the three distinct macro-states.

The three fractions of atoms are time-dependent and change during the time the 
atoms spend on the surface. The time dependence is governed by the  system of 
rate equations \cite{paper1} 
\numparts
\begin{eqnarray}
 \dot{N}_{Q} &=& - (T_{TQ}+T_{CQ})N_{Q} +  T_{QT} N_{T},\label{eq1}\\
 \dot{N}_{T} &=& -  (T_{QT}+T_{CT}) N_{T} +  T_{TQ} N_{Q},\\
 \dot{N}_{C} &=& - (\dot{N}_{Q} + \dot{N}_{T})=T_{CT}N_{T}+ T_{CQ}N_{Q} \, ,
 \label{eq:REQ}
\end{eqnarray}
\endnumparts
that is, in fact, just an example of the master equation discussed in Sec.~\ref{ss:valid-me}, and the transition rates $T_{\alpha \beta}$ 
$[\alpha, \beta = \{C, Q, T\}]$ are just the coefficients $\Gamma_{\alpha \to 
\beta}$ occurring in the latter.

In the terms used in Tab.~\ref{tab1}, this system of rate equations corresponds 
to the methods listed for the hydrodynamic stage (stage III). But, nothing 
prevents us 
 of course from using these equations also for earlier time scales, i.e., for 
the kinetic and initial stages (stage II and I, respectively).
As discussed in Sec.~\ref{ss:valid-me}, to be applicable at earlier times, we have to permit a dependence of the rates on time and on the individual probabilities 
in that case, i.e., $T_{\alpha \beta}=T_{\alpha \beta}(N_C, N_Q, N_T;t)$. 
Furthermore,  
it was shown in Ref.~\cite{paper1} that 
these rates  depend sensitively on the energy distribution function of the gas 
atoms 
$F(E^\perp, E^\parallel;t)$,  
which themselves evolve with the time duration atoms spend on the surface before 
they are
desorbed. 

In 
Ref.~\cite{paper1} 
it was demonstrated in detail, how the complete information from the MD trajectories 
can be used to explicitly reconstruct the rates $T_{\alpha \beta}$ for all 
times.
This means that 
the conditions B and C of Sec.~\ref{ss:valid-me} are also realized. 
Furthermore, the analysis revealed that the gas atom--surface interaction 
proceeds in two stages. At first, the energy distribution functions thermalize 
within the relaxation time $t_{\rm rel} \approx (20 \dots 40)$ps, which depends 
on the energy and angle of incidence of the atom. 
During this period of time the transition rates also reach their equilibrium 
form, 
$T_{\alpha \beta}(t) \to T^{\rm EQ}_{\alpha \beta}$, and remain practically 
constant for times $t > t_{\rm rel}$. This can be seen 
in~figure~\ref{fig:alex}b. 
This first time interval corresponds to the stages I and II in 
table~\ref{tab1}. 
The subsequent temporal evolution corresponds to stage III and turns out to be 
accurately described by the rate equations (\ref{eq1})-(\ref{eq:REQ}) with 
constant transition rates. 

This behavior is verified by a comparison of MD simulation data with 
the analytical solutions of the rate equations (\ref{eq1})-(\ref{eq:REQ}). The 
corresponding results are shown in figure~\ref{fig:alex}c. 
While the analytical results differ qualitatively from the MD simulation results, for short times, 
the analytical fractions of trapped and quasi-trapped atoms practically coincide 
with the MD data, for times larger than about $5t_0\approx 33$ps. This is approximately the time, where the transition rates have saturated (Fig.~\ref{fig:alex}b).

In other words,  the set of three occupation probabilities (fractions) $N_C, 
N_Q$ and $N_T$ is sufficient to capture the entire sticking and desorption 
properties of the gas atoms, 
 for times $t\ge t_{\rm rel}$. These three collective variables have emerged 
dynamically during the temporal evolution and the associated coarse graining 
dynamics. 
 Thus,  the system of rate equations is sufficient to describe the mean 
adsorption/desorption dynamics for a sufficiently large ensemble of atoms 
 at longer times. 
 The solution of these equations is computationally cheap and allows one to 
propagate the system, in principle, to arbitrarily long times. This could be of relevance for  experiments where the surface properties change in time, e.g. due to an AC field or in the course of continuous sputter deposition.
Note that the transition rates $T^{\rm EQ}_{\alpha \beta}$ obtained from the MD 
simulations are not approximations but have 
first principle quality, in 
principle.   
%

\section{Conclusion and outlook}\label{s:conclusion1}
The dramatic increase of computation power holds high promises for an improved 
simulation of plasma-surface interaction processes. This has the potential for 
major advances of this field because most current models are phenomenological using surface coefficients that are poorly known both experimentally and theoretically. Moreover, these parameter--even if they exist--may carry an (unknown) dependence on the surface conditions or the plasma parameters.

Here, we discussed the application of 
first principle
simulations to 
plasma-surface interaction where we concentrated on semiclassical molecular 
dynamics simulations where electronic degrees of freedom are not explicitly 
resolved. This is well justified for processes involving neutral particles of 
low energy where excitation or ionization can be neglected. Such MD simulations 
using accurate force fields as an input have been successfully used for more 
than two decades. 
However, the required small time steps (of the order of one femtosecond) 
prohibits, in most cases, to achieve time scales for which experimental data are available or which are of relevance for low-temperature plasma experiments.

In this paper we discussed strategies how to overcome this limitation, cf. Fig.~\ref{fig:scheme}. As a 
first concept, we briefly reviewed acceleration techniques and presented one 
recent example---\textit{Selective Process Acceleration (SPA)}~\cite{abraham_jap_16}---which is capable of achieving a boost factor of more than $10^9$. This was applied to cluster growth on a polymer substrate during sputter deposition. It was demonstrated that a controlled increase of the deposition and diffusion rates, such that ratio remains constant and on the level of the experiment, allows for a remarkable acceleration without loss of accuracy, for times up to about four minutes. These simulations can be an important piece of future plasma-surface simulations, using plasma data for the deposition rates [the flux $\textbf{J}_A^p$ in Fig. \ref{fig:theory}], as an input and delivering the time-dependent modifications of the surface morphology [labeled ``SM'' in Fig. \ref{fig:theory}] as an output for surface physics simulations.

Our second 
and main focus was on coarse graining techniques that attempt to combine two (or 
more) descriptions of different spatial and temporal resolution. Such concepts 
have existed for many years in physics, chemistry, material science and 
technology and are often summarized under the headline 
\textit{multiscale modeling}. For example, recent progress in the field of 
chemistry has been reviewed in the Nobel lectures of Levitt, Karplus and Warshel 
who shared the Nobel prize in chemistry in 2013. 

The method we have been advocating in this paper---\textit{Dynamical Freeze out of 
Dominant Modes} (DFDM)---concentrates on the idea of extending 
first principle MD simulations to long time scales without loss of accuracy. This method 
requires the derivation of  model equations that are formally exact, 
at sufficiently long time scales. Fortunately, generations of researchers have 
provided us with ample candidates for such equations which include kinetic 
equations, hydrodynamic equations, rate equations or a master equation. 
An (incomplete) overview was presented in table~\ref{tab1}. 
While these models are traditionally being used within certain approximation schemes for the relevant parameters, here 
we suggest to avoid any approximation. Instead we suggest to use \textit{exact input data} for the relevant transport coefficients or transition rates and 
to provide them by MD simulations. 

The idea of DFDM was demonstrated for the example of atom scattering from a 
metal surface, Sec.~\ref{ss:md-re}. 
It was shown that the use of MD input data in a system of rate equations 
allows to describe the sticking dynamics practically exactly. The 
first principle
MD solutions go over smoothly into the result of the rate equations 
which can be extended to macroscopic time scales. The main requirement for our 
approach to be feasible is that the time scale after which the model is valid is 
short enough to be accessible by MD simulations. 
In the case of atom sticking this time is the relaxation time $t_{\rm rel}$.  Recalling again the overview given in Fig.~\ref{fig:theory}, these simulations are capable of using the fluxes $\textbf{J}^p_A$, of atoms as an input from a plasma simulation and to return, as an output, the energy or momentum resolved fluxes $\textbf{J}^s_A$ of atoms that leave the surface.

There are various ways how to extend the present idea. If the surface is 
inhomogeneous,  a straightforward generalization would be to include the space 
dependence into the densities and the rates. Then, the rate equations turn into 
hydrodynamic equations. Furthermore, the effect of a plasma environment, such as 
characteristic particle fluxes or an adsorbate-covered surface, are 
straightforwardly included into our scheme, as discussed in Ref.~\cite{paper1}.

To go beyond the problem of neutral atom sticking, the present theoretical idea can be straightforwardly extended also to other analytical models,  
which are listed in table~\ref{tab1}. One example are hydrodynamic equations for the particle densities and fluxes, e.g. in the plasma bulk or in the sheath. Instead of using an approximate decoupling, e.g. by using a model equation of state, one can  use MD data as an input again. 
Finally, the idea of DFDM can also be extended to quantum systems where the 
dynamics are treated by quantum molecular dynamics or time-dependent DFT. These 
\textit{ab initio} input data can again be linked to macroscopic model equations 
such as diffusion equation, as in Ref.~\cite{franke_prb_10}, hydrodynamic equations or quantum hydrodynamics, e.g. \cite{michta_cpp15, zhandos_pop18}.

Finally, let us return to the overview on theoretical methods for the plasma--solid interface that was sketched in Fig.~\ref{fig:theory}. While, in this paper, we have concentrated on molecular dynamics, the box in the center indicates that there is a much broader arsenal of tools available. In fact there is no unique method that allows to describe all processes.
In particular for the description of electrons and ions crossing the interface, semiclassical MD fails, and quantum approaches are necessary. This concerns the neutralization of low-energy ions, e.g. \cite{pamperin_prb15} and their stopping in the solid, as well as the electron dynamics across the interface, e.g. \cite{bronold_prl15}. Here nonequlibrium quantum methods such as time-dependent DFT or nonequlibrium Green functions simulations, e.g. \cite{zhao_prb_15, balzer_prb_16, balzer_lnp_13} have to be used. These methods are extremely expensive, and the goal will have to be to use their results as input to simpler approaches such as quantum kinetic equations \cite{bonitz_qkt} or improved molecular dynamics simulations. Moreover, to properly capture the influence of the plasma on the solid, these surface simulations have to be linked to fluid or kinetic simulations of the plasma, as indicated by the arrows in Fig.~\ref{fig:theory}.

This connection maybe summarized by adding a fifth step to the list of Sec.~\ref{s:intro},  meaning that plasma simulations will have to be supplied with accurate fluxes of electrons, ions and neutrals leaving the surface and, at the same time, provide those fluxes that impact the solid, to surface simulations. As a result of the fluxes across the interface, the specific plasma conditions are expected to influence the surface properties, such as surface roughness, morphology or chemical reactivity.   Ultimately, an integrated modeling of the plasma and the solid surface will be required \cite{interface} to overcome the trial and error character of many experiments and to achieve a predictive modeling of the relevant processes.


\section*{Acknowledgements}
We acknowledge discussions with E. Pehlke und B. Hartke.



\begin{thebibliography}{60}
\bibitem{comment_exactness} This means we consider here only models that have 
this property.
\end{thebibliography}
\end{document}